\RequirePackage{tikz}

\documentclass[sn-basic]{sn-jnl}
\jyear{2021}

\usepackage{natbib}
\usepackage{subfigure}
\usepackage{bm}
\usepackage{booktabs}

\usepackage{tikz}
\usetikzlibrary{shapes.misc}
\usetikzlibrary{matrix}
\usetikzlibrary{arrows,backgrounds,fit,calc,shapes,automata}
\usetikzlibrary{bayesnet}
\tikzstyle{every picture}+=[remember picture]
\tikzset{
    >=stealth',
    punkt/.style={
           rectangle,
           rounded corners,
           draw=black,
           thick,
           text width=3.5cm,
           minimum height=1.0cm,
           text centered},
    pil/.style={
           ->,
           thick,
           shorten <=2pt,
           shorten >=2pt,}
}
\usetikzlibrary{decorations.pathreplacing,shapes}

\begin{document}

%% *** Frontmatter ***

%===============================================================================

%==============================================================================
% Comments
%==============================================================================

\newcommand{\todo}[1]{\textcolor{red}{[{\bf TODO}: #1]}}
\newcommand{\jdm}[1]{\textcolor{red}{[{\bf JDM}: #1]}}

%==============================================================================
% Abbreviations
%==============================================================================

\newcommand{\eqn}[1]{(#1)}
\newcommand{\Eqn}[1]{(#1)}
\newcommand{\tbl}[1]{Table~#1}
\newcommand{\Tbl}[1]{Table~#1}
\newcommand{\fig}[1]{Fig.~#1}
\newcommand{\Fig}[1]{Fig.~#1}
\newcommand{\sectn}[1]{Sec.~#1}
\newcommand{\Sectn}[1]{Sec.~#1}
\newcommand{\chap}[1]{Chapter~#1}
\newcommand{\Chap}[1]{Chapter~#1}
\newcommand{\appn}[1]{Appendix~#1}
\newcommand{\Appn}[1]{Appendix~#1}
\newcommand{\lemmaref}[1]{Lemma~#1}

\newcommand{\etal}{\mbox{\it et al.}}
\newcommand{\eg}{\mbox{\it e.g.}}
\newcommand{\ie}{\mbox{\it i.e.}}
\newcommand{\etc}{\mbox{\it etc.}}
\newcommand{\cf}{\mbox{\it cf.}}

%==============================================================================
% Units
%==============================================================================

\newcommand{\ghz}{{GHz}}
\newcommand{\joule}{{J}}
\newcommand{\watt}{{W}}
\newcommand{\metre}{m}
\newcommand{\nanometre}{nm}
\newcommand{\steradian}{sr}
\newcommand{\kelvin}{{K}}
\newcommand{\degrees}{\ensuremath{{^\circ}}}

%==============================================================================
% Acronyms
%=============================================================================

\newcommand{\fft}{{FFT}}
\newcommand{\dft}{{DFT}}

\newcommand{\frw}{{FRW}}
\newcommand{\frwtext}{{Friedmann-Robertson-Walker}}

\newcommand{\cmb}{{CMB}}
\newcommand{\cmbtext}{{cosmic microwave background}}
\newcommand{\Cmbtext}{{Cosmic microwave background}}

\newcommand{\wmap}{{WMAP}}
\newcommand{\wmaptext}{{Wilkinson Microwave Anisotropy Probe}}
\newcommand{\ilc}{{ILC1}}
\newcommand{\ilcthree}{{ILC3}}
\newcommand{\ilctext}{{internal linear combination}}

\newcommand{\planck}{\textit{Planck}}

\newcommand{\cobedmr}{\mbox{COBE-DMR}}
\newcommand{\cobedmrtext}{Cosmic Background Explorer-Differential Microwave Radiometer}
\newcommand{\cobe}{\mbox{COBE}}

\newcommand{\bianchi}{{Bianchi}}
\newcommand{\bianchiviih}{{Bianchi VII{$_{\lowercase{h}}$}}}
\newcommand{\wmapbianchi}{{\wmap1$-$Bianchi}}

\newcommand{\isw}{{ISW}}
\newcommand{\iswtext}{{integrated Sachs-Wolfe}}
\newcommand{\Iswtext}{{Integrated Sachs-Wolfe}}

\newcommand{\cswt}{{CSWT}}
\newcommand{\cswti}{\mbox{CSWT$_{\rm I}$}}
\newcommand{\cswtii}{\mbox{CSWT$_{\rm II}$}}
\newcommand{\cswttext}{continuous spherical wavelet transform}
\newcommand{\Cswttext}{Continuous spherical wavelet transform}

\newcommand{\lss}{{LSS}}
\newcommand{\lsstext}{large scale structure}
\newcommand{\Lsstext}{Large scale structure}

\newcommand{\nvss}{{NVSS}}
\newcommand{\nvsstext}{{NRAO VLA Sky Survey}}

\newcommand{\sdss}{{SDSS}}
\newcommand{\sdsstext}{Sloan Digital Sky Survey}

\newcommand{\twomass}{{2MASS}}
\newcommand{\twomasstext}{Two Micron All Sky Survey}

\newcommand{\lcdm}{\ensuremath{\Lambda}{CDM}}
\newcommand{\lcdmtext}{\ensuremath{\Lambda} Cold Dark Matter}

\newcommand{\ska}{{SKA}}
\newcommand{\skatext}{Square Kilometre Array}

\newcommand{\smhwtext}{{spherical Mexican hat wavelet}}
\newcommand{\smhw}{{SMHW}}
\newcommand{\semhw}{{SEMHW}}
\newcommand{\smw}{{SMW}}
\newcommand{\sbwtext}{{spherical butterfly wavelet}}
\newcommand{\sbw}{{SBW}}
\newcommand{\mexhat}{Mexican hat}
\newcommand{\Mexhat}{Mexican hat}
\newcommand{\morlet}{real Morlet}
\newcommand{\Morlet}{Real Morlet}

\newcommand{\healpix}{{\tt HEALPix}}
\newcommand{\healpixtext}{{Hierarchical Equal Area isoLatitude Pixelisation of the sphere}}
\newcommand{\cmbfast}{\mbox{\tt CMBFAST}}
\newcommand{\camb}{\mbox{\tt CAMB}}
\newcommand{\cosmomc}{\mbox{\tt COSMOMC}}
\newcommand{\anicosmo}{\mbox{\tt ANICOSMO}}

\newcommand{\yawtbtext}{{Yet Another Wavelet Toolbox}}
\newcommand{\yawtb}{{\tt YAWTb}}

\newcommand{\stwo}{{\tt S2}}
\newcommand{\stwofil}{{\tt S2FIL}}
\newcommand{\fastcswt}{{\tt FastCSWT}}
\newcommand{\comb}{{\tt COMB}}
\newcommand{\bianchicode}{{\tt Bianchi}}
\newcommand{\bianchicodetwo}{{\tt Bianchi2}}
\newcommand{\sshtcode}{{\tt SSHT}}
\newcommand{\sothreecode}{{\tt SO3}}
\newcommand{\stwoletcode}{{\tt S2LET}}

\newcommand{\lambdaarch}{{LAMBDA}}
\newcommand{\lambdaarchtext}{{Legacy Archive for Microwave Background Data Analysis}}

\newcommand{\kpzero}{{Kp0}}
\newcommand{\tohmap}{{TOH}}

\newcommand{\fwhm}{{FWHM}}
\newcommand{\fwhmtext}{{full-width-half-maximum}}
\newcommand{\snr}{{\rm SNR}}
\newcommand{\snrp}{\ensuremath{\snr_{\pind}}}
\newcommand{\snrs}{\ensuremath{\snr_{\sind}}}

\newcommand{\mf}{{MF}}
\newcommand{\saf}{{SAF}}

\newcommand{\fov}{FOV}
\newcommand{\wfov}{WFOV}
\newcommand{\wterm}{\mbox{$w$-term}}

%==============================================================================
% Data
%==============================================================================

\newcommand{\spotloc}{\mbox{\ensuremath{(l,b)=(209^\circ,-57^\circ)}}}

%==============================================================================
% Maths
%==============================================================================

% General
\newcommand{\spcend}{\ensuremath{\:}}
\newcommand{\img}{\ensuremath{{\rm i}}}
\newcommand{\cconj}{\ensuremath{\ast}} 
\newcommand{\adjoint}{\ensuremath{{}^\dagger}} 
\newcommand{\reals}{\ensuremath{\mathbb{R}}}
\newcommand{\realsnn}{\ensuremath{\mathbb{R^+}}}
\newcommand{\realsnz}{\ensuremath{\mathbb{R}^{+}_{\ast}}}
\newcommand{\integers}{\ensuremath{\mathbb{Z}}}
\newcommand{\naturals}{\ensuremath{\mathbb{N}}}
\newcommand{\complex}{\ensuremath{\mathbb{C}}}
\newcommand{\ltwo}{\ensuremath{\mathrm{L}^2}}
\newcommand{\sphere}{\ensuremath{{\mathbb{S}^2}}}
\newcommand{\sothree}{\ensuremath{{\mathrm{SO}(3)}}}
\newcommand{\sotwo}{\ensuremath{{\mathrm{SO}(2)}}}
\newcommand{\torus}{\ensuremath{{\mathrm{T}^2}}}
\newcommand{\disk}{\ensuremath{{D^2}}}
\newcommand{\ball}{{\ensuremath{\mathbb{B}^3}}}
\newcommand{\ipball}{\ensuremath{\ball}}
\newcommand{\ipsphere}{\ensuremath{\sphere}}
\newcommand{\iphalfline}{\ensuremath{\realsnn}}
\newcommand{\vect}[1]{\ensuremath{\mbox{\boldmath ${#1}$}}}
\newcommand{\vectsm}[1]{\ensuremath{\mbox{\boldmath \footnotesize ${#1}$}}}
\newcommand{\xvect}{\vect{x}}
\newcommand{\kvect}{\vect{k}}
\newcommand{\zerovect}{\ensuremath{\mathbf{0}}}
\newcommand{\knaut}{\ensuremath{k_0}}
\newcommand{\nsigma}{\ensuremath{N_\sigma}}
\newcommand{\zreal}{{\ensuremath{\rm{Re}}}}
\newcommand{\zimag}{{\ensuremath{\rm{Im}}}}
\newcommand{\dotprod}{\ensuremath{\cdot}}
\newcommand{\likelihood}{\ensuremath{\mathcal{L}}}
\newcommand{\prior}{\ensuremath{\pi}}
\newcommand{\opnexpv}{\ensuremath{\langle}}
\newcommand{\clsexpv}{\ensuremath{\rangle}}

% Integrals
\newcommand{\dx}{\ensuremath{\mathrm{\,d}}}
\newcommand{\dmu}[1]{\ensuremath{\dx \Omega(#1)}}
\newcommand{\dmun}{\ensuremath{\dx \Omega}}
\newcommand{\deul}[1]{\ensuremath{\dx \varrho(#1)}}
\newcommand{\deuln}{\ensuremath{\dx \varrho}}
\newcommand{\ddr}{\ensuremath{\frac{\partial}{\partial\scale}}}
\newcommand{\innerp}[2]{\ensuremath{\langle {#1},\: {#2} \rangle}}

% Theorems
% \theoremstyle{remark}
% \newtheorem{lemma}{Lemma}[section]
% \theoremstyle{plain}
% \newtheorem{result}{Result}
% \theoremstyle{definition}
% \newtheorem*{definition}{Definition}
% \newcommand{\thmend}{{\mbox{}  \hfill \raggedright \ensuremath{\blacksquare}\\[3mm]}}

% Physics of the universe
\newcommand{\clight}{\ensuremath{c}}
\newcommand{\ctime}{\ensuremath{\eta}}
\newcommand{\ptime}{\ensuremath{t}}
\newcommand{\pdist}{\ensuremath{d}}
\newcommand{\rvel}{\ensuremath{v}}
\newcommand{\hub}{\ensuremath{H}}
\newcommand{\hubsmall}{\ensuremath{h}}
\newcommand{\scfac}{\ensuremath{R}}
\newcommand{\scfacd}{\ensuremath{\dot{\scfac}}}
\newcommand{\scfacdd}{\ensuremath{\ddot{\scfac}}}
\newcommand{\kcurv}{\ensuremath{k}}
\newcommand{\cconst}{\ensuremath{\Lambda}}
\newcommand{\gconst}{\ensuremath{G}}
\newcommand{\den}{\ensuremath{\rho}}
\newcommand{\denmat}{\ensuremath{\rho_{\rm m}}}
\newcommand{\denrad}{\ensuremath{\rho_{\rm r}}}
\newcommand{\dencrit}{\ensuremath{\rho_{\rm c}}}
\newcommand{\Den}{\ensuremath{\Omega}}
\newcommand{\Dentot}{\ensuremath{\Den_{\rm total}}}
\newcommand{\Denmat}{\ensuremath{\Den_{\rm m}}}
\newcommand{\Denmatc}{\ensuremath{\Den_{\rm c}}}
\newcommand{\Denmatb}{\ensuremath{\Den_{\rm b}}}
\newcommand{\Dencurvature}{\ensuremath{\Den_{k}}}
\newcommand{\Denrad}{\ensuremath{\Den_{\rm r}}}
\newcommand{\Denlambda}{\ensuremath{\Den_{\Lambda}}}
\newcommand{\pres}{\ensuremath{p}}
\newcommand{\zrec}{\ensuremath{\z_{\rm rec}}}

% CMB
\newcommand{\dtemp}{\ensuremath{\Delta T}}
\newcommand{\temp}{\ensuremath{T_0}}
\newcommand{\rdtemp}{\ensuremath{\frac{\dtemp}{\temp}}}
\newcommand{\gravpotent}{\ensuremath{\delta\Phi}}
\newcommand{\cl}{\ensuremath{C_\el}}
\newcommand{\clest}{\ensuremath{\widehat{C}_\el}}

% Variables (e.g. spherical coordinates, indices, scales)
\newcommand{\sa}{\ensuremath{\omega}}
\newcommand{\saa}{\ensuremath{\theta}}
\newcommand{\sab}{\ensuremath{\varphi}}
\newcommand{\sas}{\ensuremath{\saa, \sab}}
\newcommand{\radvec}{\ensuremath{\vect{\rad}}}
\newcommand{\radvecunit}{\ensuremath{\hat{\vect{\rad}}}}
\newcommand{\eul}{\ensuremath{\mathbf{\rho}}}
\newcommand{\euls}{\ensuremath{\eula, \eulb, \eulc}}
\newcommand{\eula}{\ensuremath{\alpha}}
\newcommand{\eulb}{\ensuremath{\beta}}
\newcommand{\eulc}{\ensuremath{\gamma}}
\newcommand{\eulai}{\ensuremath{a}}
\newcommand{\eulbi}{\ensuremath{b}}
\newcommand{\eulci}{\ensuremath{g}}
\newcommand{\eulaiang}{\ensuremath{\eula_\eulai}}
\newcommand{\eulbiang}{\ensuremath{\eulb_\eulbi}}
\newcommand{\eulciang}{\ensuremath{\eulc_\eulci}}
\newcommand{\el}{\ensuremath{\ell}}
\newcommand{\m}{\ensuremath{m}}
\newcommand{\n}{\ensuremath{n}}
\newcommand{\ip}{\ensuremath{p}}
\newcommand{\spin}{\ensuremath{s}}
\newcommand{\elm}{\ensuremath{{\el\m}}}
\newcommand{\elp}{\ensuremath{{\el\p}}}
\newcommand{\elmax}{\ensuremath{{L}}}
\newcommand{\mmax}{\ensuremath{{M}}}
\newcommand{\nmax}{\ensuremath{{N}}}
\newcommand{\ipmax}{\ensuremath{{P}}}
\newcommand{\pp}{\ensuremath{^{\prime\prime}}}
\newcommand{\scale}{\ensuremath{R}}
\newcommand{\scalenaut}{\ensuremath{R_0}}
\newcommand{\scalea}{\ensuremath{a}}
\newcommand{\scaleb}{\ensuremath{b}}
\newcommand{\scaleab}{\ensuremath{{\scalea,\scaleb}}}
\newcommand{\rad}{\ensuremath{r}}
\newcommand{\cmbtemp}{\ensuremath{T}}
\newcommand{\pixw}{\ensuremath{p_\el}}
\newcommand{\pind}{\ensuremath{{\rm p}}}
\newcommand{\sind}{\ensuremath{{\rm s}}}
\newcommand{\mind}{\ensuremath{{m}}}
\newcommand{\tfov}{\ensuremath{\saa_{\rm \fov}}}

% Coordinate systems
\newcommand{\nind}{\ensuremath{\hat{\vect{n}}_i}}
\newcommand{\none}{\ensuremath{\hat{\vect{n}}_1}}
\newcommand{\ntwo}{\ensuremath{\hat{\vect{n}}_2}}
\newcommand{\nthree}{\ensuremath{\hat{\vect{n}}_3}}
\newcommand{\gind}{\ensuremath{\hat{\vect{g}}_i}}
\newcommand{\gone}{\ensuremath{\hat{\vect{g}}_1}}
\newcommand{\gtwo}{\ensuremath{\hat{\vect{g}}_2}}
\newcommand{\gthree}{\ensuremath{\hat{\vect{g}}_3}}
\newcommand{\eind}{\ensuremath{\hat{\vect{e}}_i}}
\newcommand{\eone}{\ensuremath{\hat{\vect{e}}_1}}
\newcommand{\etwo}{\ensuremath{\hat{\vect{e}}_2}}
\newcommand{\ethree}{\ensuremath{\hat{\vect{e}}_3}}
\newcommand{\uone}{\ensuremath{\hat{\vect{u}}_1}}
\newcommand{\utwo}{\ensuremath{\hat{\vect{u}}_2}}
\newcommand{\uthree}{\ensuremath{\hat{\vect{u}}_3}}

% Special functions and harmonic coefficients
\newcommand{\kron}[2]{\ensuremath{\delta_{{#1}{#2}}}}
\newcommand{\kronsp}[2]{\ensuremath{\delta_{{#1},{#2}}}}
\newcommand{\dirac}{\ensuremath{\delta}}
\newcommand{\poly}{\ensuremath{\phi}}
\newcommand{\polyw}{\ensuremath{w}}
\renewcommand{\exp}[1]{\ensuremath{{\rm e}^{#1}}}
\newcommand{\shfarg}[3]{\ensuremath{Y_{#1#2}({#3})}}
\newcommand{\shfargc}[3]{\ensuremath{Y_{#1#2}^\cconj({#3})}}
\newcommand{\shfargsp}[3]{\ensuremath{Y_{{#1},{#2}}({#3})}}
\newcommand{\sshfarg}[4]{\ensuremath{{{}_{#4} Y_{#1#2}({#3})}}}
\newcommand{\sshfargc}[4]{\ensuremath{{{}_{#4} Y_{#1#2}^\cconj({#3})}}}
\newcommand{\sshfargsp}[4]{\ensuremath{{{}_{#4} Y_{{#1},{#2}}({#3})}}}
\newcommand{\shf}[2]{\ensuremath{Y_{#1#2}}}
\newcommand{\shfc}[2]{\ensuremath{Y_{#1#2}^\cconj}}
\newcommand{\shc}[3]{\ensuremath{{#1}_{{#2}{#3}}}}
\newcommand{\shcc}[3]{\ensuremath{{#1}_{{#2}{#3}}^\cconj}}
\newcommand{\shcsp}[3]{\ensuremath{{#1}_{{#2},{#3}}}}
\newcommand{\shccsp}[3]{\ensuremath{{#1}_{{#2},{#3}}^\cconj}}
\newcommand{\sshf}[3]{\ensuremath{{\prescript{}{#3} Y_{#1#2}}}}
\newcommand{\sshfc}[3]{\ensuremath{{\prescript{}{#3} Y_{#1#2}^\cconj}}}
\newcommand{\sshc}[4]{\ensuremath{\prescript{}{#4} {#1}_{{#2}{#3}}}}
\newcommand{\sshcc}[4]{\ensuremath{\prescript{}{#4} {#1}_{{#2}{#3}}^\cconj}}
\newcommand{\sshcsp}[4]{\ensuremath{\prescript{}{#4} {#1}_{{#2},{#3}}}}
\newcommand{\sshccsp}[4]{\ensuremath{\prescript{}{#4} {#1}_{{#2},{#3}}^\cconj}}
\newcommand{\leg}[2]{\ensuremath{P_{{#1}}({#2})}}
\newcommand{\aleg}[3]{\ensuremath{P_{#1}^{#2}({#3})}}
\newcommand{\legc}[2]{\ensuremath{{#1}_{#2}}}
\newcommand{\daleg}[3]{\ensuremath{P_{#1}^{#2\prime}({#3})}}
\newcommand{\sbessel}[1]{\ensuremath{j_{#1}}}
\newcommand{\bessel}[2]{\ensuremath{J_{#1}({#2})}}
\newcommand{\jacobi}[4]{\ensuremath{P_{#1}^{(#2,#3)}({#4})}}
\newcommand{\jacobia}{\ensuremath{a}}
\newcommand{\jacobib}{\ensuremath{b}}
\newcommand{\gammafun}{\ensuremath{\Gamma}}
\newcommand{\binomial}[2]{\ensuremath{\,\, {}^{#1} C_{#2} \,\,}}

% Wigner functions
\newcommand{\dmatbig}{\ensuremath{D}}
\newcommand{\Dlmn}{\ensuremath{ \dmatbig_{\m\n}^{\el} }}
\newcommand{\Dlmnc}{\ensuremath{ \dmatbig_{\m\n}^{\el\cconj} }}
\newcommand{\Dlmz}{\ensuremath{ \dmatbig_{\m0}^{\el} }}
\newcommand{\Dlmzc}{\ensuremath{ \dmatbig_{\m0}^{\el\cconj} }}
\newcommand{\Dlmnp}{\ensuremath{ \dmatbig_{\m\n}^{\el}(\eul) }}
\newcommand{\Dlmnpc}{\ensuremath{ \dmatbig_{\m\n}^{\el\cconj}(\eul) }}
\newcommand{\Dlmnabg}{\ensuremath{ \dmatbig_{\m\n}^{\el}(\euls) }}
\newcommand{\Dlmnabgc}{\ensuremath{ \dmatbig_{\m\n}^{\el\cconj}(\euls) }}
\newcommand{\dmatsmall}{\ensuremath{d}}
\newcommand{\dlmn}{\ensuremath{ \dmatsmall_{\m\n}^{\el} }}
\newcommand{\dlmnc}{\ensuremath{ \dmatsmall_{\m\n}^{\el\cconj} }}
\newcommand{\dlnm}{\ensuremath{ \dmatsmall_{\n\m}^{\el} }}
\newcommand{\dlmz}{\ensuremath{ \dmatsmall_{\m0}^{\el} }}
\newcommand{\dlmzc}{\ensuremath{ \dmatsmall_{\m0}^{\el\cconj} }}
\newcommand{\dlmnb}{\ensuremath{ \dmatsmall_{\m\n}^{\el}(\eulb) }}
\newcommand{\dlmnbc}{\ensuremath{ \dmatsmall_{\m\n}^{\el\cconj}(\eulb) }}
\newcommand{\dlnmb}{\ensuremath{ \dmatsmall_{\n\m}^{\el}(\eulb) }}
\newcommand{\dlmnhalfpi}[3]{\ensuremath{ \Delta_{{#2}{#3}}^{#1} }}
\newcommand{\dlmnhalfpic}[3]{\ensuremath{ \Delta_{{#2},{#3}}^{#1} }}
\newcommand{\dlmnhalfpim}{\ensuremath{ \Delta_{{\m\p}{\m}}^{\el} }}
\newcommand{\dlmnhalfpin}{\ensuremath{ \Delta_{{\m\p}{\n}}^{\el} }}
\newcommand{\dlmnhalfpisn}{\ensuremath{ \Delta_{{\m\p}{,-\spin}}^{\el} }}
\newcommand{\wigc}[4]{\ensuremath{{#1}^{#2}_{{#3}{#4}}}}

% Operators (e.g. rotations and dilations)
\newcommand{\dil}{\ensuremath{\mathcal{D}}}
\newcommand{\dilsmall}{\ensuremath{d}}
\newcommand{\spo}{\ensuremath{\Pi}}
\newcommand{\rot}{\ensuremath{\mathcal{R}}}
\newcommand{\rotarg}[1]{\ensuremath{\mathcal{R}_{#1}}}
\newcommand{\rotmat}{\ensuremath{\mathbf{\mathsf{R}}}}
\newcommand{\rotmatarg}[1]{\ensuremath{\rotmat_{#1}}}
\newcommand{\transl}{\ensuremath{\mathcal{T}}}
\newcommand{\spinup}{\ensuremath{\eth}}
\newcommand{\spindown}{\ensuremath{\bar{\eth}}}

% Spherical wavelets
\newcommand{\sky}{\ensuremath{f}}
\newcommand{\skywav}{\ensuremath{{\mathcal{W}_\wav^\sky}}}
\newcommand{\skywavni}{\ensuremath{{\mathcal{W}}}}
\newcommand{\skywavi}{\ensuremath{{\widehat{\mathcal{W}}_\wavi^\sky}}}
\newcommand{\skyLwavii}{\ensuremath{{\widetilde{\mathcal{W}}_\wavm^{\Lopii\sky}}}}
\newcommand{\skywavii}{\ensuremath{{\widetilde{\mathcal{W}}_\wavm^\sky}}}
\newcommand{\wavop}{\ensuremath{\mathcal{W}}}
\newcommand{\wav}{\ensuremath{\psi}}
\newcommand{\swav}{\ensuremath{{}_\spin\psi}}
\newcommand{\wavi}{\ensuremath{\Phi}}
\newcommand{\wavii}{\ensuremath{\Psi}}
\newcommand{\wavm}{\ensuremath{\Upsilon}}
\newcommand{\wavmc}{\ensuremath{\Upsilon^\cconj}}
\newcommand{\wavs}{\ensuremath{\Phi}}
\newcommand{\swavs}{\ensuremath{{}_\spin\Phi}}
\newcommand{\wcoeff}{\ensuremath{W}}
\newcommand{\scoeff}{\ensuremath{W}}
\newcommand{\wscale}{\ensuremath{j}}
\newcommand{\wscalemax}{\ensuremath{J}}
\newcommand{\wscalemin}{\ensuremath{J_0}}
\newcommand{\wposn}{\ensuremath{\eul}}
\newcommand{\wscaleposn}{\ensuremath{{\wscale\wposn}}}
\newcommand{\dilparam}{\ensuremath{\alpha}}
\newcommand{\wavker}{\ensuremath{\kappa}}
\newcommand{\wavsteer}{\ensuremath{\zeta}}
\newcommand{\steerinterp}{\ensuremath{z}}
\newcommand{\admissC}{\ensuremath{C}}
\newcommand{\admissCi}{\ensuremath{\widehat{\admissC}_\wavi}}
\newcommand{\admissCli}{\ensuremath{\widehat{\admissC}_\wavi^\el}}
\newcommand{\admissCii}{\ensuremath{\widetilde{\admissC}_\wavm}}
\newcommand{\admissClii}{\ensuremath{\widetilde{\admissC}_\wavm^\el}}
\newcommand{\Lopi}{\ensuremath{\widehat{L}_\wavi}}
\newcommand{\Lopii}{\ensuremath{\widetilde{L}_\wavm}}
\newcommand{\cocycle}{\ensuremath{\lambda}}
\newcommand{\effsize}{\ensuremath{\xi}}
\newcommand{\eccen}{\ensuremath{\epsilon}}
\newcommand{\sigx}{\ensuremath{\sigma_x}}
\newcommand{\sigy}{\ensuremath{\sigma_y}}

% Radon and ridgelet transforms
\newcommand{\sradon}{\ensuremath{\mathcal{S}}}
\newcommand{\sradondelta}{\ensuremath{\xi}}
\newcommand{\rwav}{\ensuremath{\psi}}
\newcommand{\srwav}{\ensuremath{{}_\spin\psi}}
\newcommand{\rwavs}{\ensuremath{\phi}}
\newcommand{\srwavs}{\ensuremath{{}_\spin\phi}}

\newcommand{\rwcoeff}{\ensuremath{G}}
\newcommand{\rscoeff}{\ensuremath{R}}
\newcommand{\rwavop}{\ensuremath{\mathcal{G}}}

% Common sums and fractions
\newcommand{\elmfact}{\ensuremath{\frac{(\el-\m)!}{(\el+\m)!}}}
\newcommand{\elpifrac}{\ensuremath{\frac{2\el+1}{4\pi}}}
\newcommand{\elpifracinv}{\ensuremath{\frac{4\pi}{2\el+1}}}
\newcommand{\sumlm}{\ensuremath{\sum_{\el=0}^{\infty} \sum_{\m=-\el}^\el}}
\newcommand{\sumlmn}{\ensuremath{\sum_{\el=0}^{\infty} \sum_{\m=-\el}^\el} \sum_{\n=-\el}^\el}
\newcommand{\suml}{\ensuremath{\sum_{\el=0}^{\infty}}}
\newcommand{\sumlmbl}{\ensuremath{\sum_{\el=0}^{\elmax-1} \sum_{\m=-\el}^\el}}
\newcommand{\sumlmnbl}{\ensuremath{\sum_{\el=0}^{\elmax-1} \sum_{\m=-\el}^\el} \sum_{\n=-\el}^\el}
\newcommand{\sumlbl}{\ensuremath{\sum_{\el=0}^{\elmax-1}}}
\newcommand{\summ}{\ensuremath{\sum_{\m=-\el}^\el}}
\newcommand{\summp}{\ensuremath{\sum_{\m\p=-\el}^\el}}
\newcommand{\sumn}{\ensuremath{\sum_{\n=-\el}^\el}}
\newcommand{\sumulm}{\ensuremath{\sum_{\el\m}}}
\newcommand{\sumulmn}{\ensuremath{\sum_{\el\m\n}}}
\newcommand{\sumlmp}{{\ensuremath{\sum_{\el\p =0}^{\infty} \: \sum_{\m\p=-\el\p }^{\el\p } \:}}}
\newcommand{\sumllm}{{\ensuremath{\sum_{\el =0}^{\infty} \: \sum_{\el\p =0}^{\infty} \: \sum_{\m\p=-\el }^{\el } \:}}}
\newcommand{\sumlmb}{\ensuremath{\sum_{\el \m}}}
\newcommand{\sumlmpb}{\ensuremath{\sum_{\elp \m\p}}}
\newcommand{\nl}{\ensuremath{\sqrt{\frac{2\el+1}{4\pi}}}}
\newcommand{\nlm}{\ensuremath{\sqrt{\frac{2\el+1}{4\pi}\frac{(\el-\m)!}{(\el+\m)!}}}}
\newcommand{\nlpmp}{\ensuremath{\sqrt{\frac{2\el\p+1}{4\pi}\frac{(\el\p-\m\p)!}{(\el\p+\m\p)!}}}}
\newcommand{\nlpm}{\ensuremath{\sqrt{\frac{2\el\p+1}{4\pi}\frac{(\el\p-\m)!}{(\el\p+\m)!}}}}

% Fast spin spherical harmonic transform
\newcommand{\G}[3]{\ensuremath{{{}_{#1} G_{{#2} {#3}}}}}
\newcommand{\Gtilde}[3]{\ensuremath{{{}_{#1} \tilde{G}_{{#2} {#3}}}}}
\newcommand{\rG}[3]{\ensuremath{{{}_{#1} \tilde{G}_{{#2} {#3}}}}}
\newcommand{\Gsm}{\ensuremath{\G{\spin}{\m}{}}}
\newcommand{\Gsmm}{\ensuremath{\G{\spin}{\m}{\m\p}}}
\newcommand{\Gsmt}{\ensuremath{\G{\spin}{\m}{}(\saa)}}
\newcommand{\rGsmt}{\ensuremath{\rG{\spin}{\m}{}(\saa)}}
\newcommand{\rGsmtn}{\ensuremath{\rG{\spin}{\m}{}(-\saa)}}
\newcommand{\Gsmti}{\ensuremath{\G{\spin}{\m}{}({\saaiang})}}
\newcommand{\Gsmtin}{\ensuremath{\G{\spin}{\m}{}({-\saaiang})}}
\newcommand{\rGsmti}{\ensuremath{\rG{\spin}{\m}{}({\saaiang})}}
\newcommand{\Gmn}{\ensuremath{G_{\m\n}}}
\newcommand{\Gmnm}{\ensuremath{G_{\m\n\m\p}}}
\newcommand{\Gmnb}{\ensuremath{G_{\m\n}(\eulb)}}
\newcommand{\Gmnbi}{\ensuremath{G_{\m\n}(\eulbiang)}}
\newcommand{\rGmnb}{\ensuremath{\tilde{G}_{\m\n}(\eulb)}}
\newcommand{\rGmnbi}{\ensuremath{\tilde{G}_{\m\n}(\eulbiang)}}
\newcommand{\Fmnm}{\ensuremath{\tilde{G}_{\m\n\m\p}}}
\newcommand{\Fmnmp}{\ensuremath{\tilde{G}_{\m\n\m\p{}\p}}}
\newcommand{\FImnm}{\ensuremath{F_{\m\n\m\p}}}
\newcommand{\F}[3]{\ensuremath{{{}_{#1} F_{{#2} {#3}}}}}
\newcommand{\rF}[3]{\ensuremath{{{}_{#1} \tilde{F}_{{#2} {#3}}}}}
\newcommand{\Fsm}{\ensuremath{\F{\spin}{\m}{}}}
\newcommand{\rFsm}{\ensuremath{\rF{\spin}{\m}{}}}
\newcommand{\Fsmm}{\ensuremath{\F{\spin}{\m}{\m\p}}}
\newcommand{\Fzmm}{\ensuremath{\F{0}{\m}{\m\p}}}
\newcommand{\Fsmmn}{\ensuremath{\F{\spin}{\m}{,-\m\p}}}
\newcommand{\Fsmnmn}{\ensuremath{\F{\spin}{-\m}{,-\m\p}}}
\newcommand{\Fzmmn}{\ensuremath{\F{0}{\m}{,-\m\p}}}
\newcommand{\Fzmnmn}{\ensuremath{\F{0}{-\m}{,-\m\p}}}
\newcommand{\Fsmmp}{\ensuremath{\F{\spin}{\m}{\m{\p}{\p}}}}
\newcommand{\Fsmt}{\ensuremath{\F{\spin}{\m}{}(\saa)}}
\newcommand{\rFsmt}{\ensuremath{\rF{\spin}{\m}{}(\saa)}}
\newcommand{\Fsmtn}{\ensuremath{\F{\spin}{\m}{}(-\saa)}}
\newcommand{\Fsmpt}{\ensuremath{\F{\spin}{\m\p}{}(\saa)}}
\newcommand{\intsaa}{\ensuremath{\int_0^{\pi} \dx \saa \sin \saa}}
\newcommand{\intsab}{\ensuremath{\int_0^{2\pi} \dx \sab}}
\newcommand{\saai}{\ensuremath{t}}
\newcommand{\sabi}{\ensuremath{p}}
\newcommand{\saaiang}{\ensuremath{\saa_\saai}}
\newcommand{\sabiang}{\ensuremath{\sab_\sabi}}
\newcommand{\sais}{\ensuremath{\sabi,\saai}}
\newcommand{\saisang}{\ensuremath{\saaiang,\sabiang}}
\newcommand{\sumsaai}{\ensuremath{\sum_{\saai=-(\elmax-1)}^{\elmax-1}}}
\newcommand{\sumsabi}{\ensuremath{\sum_{\sabi=-(\elmax-1)}^{\elmax-1}}}
\newcommand{\sumltrunc}{\ensuremath{\sum_{\el=0}^{\elmax-1}}}
\newcommand{\summtrunc}{\ensuremath{\sum_{\m=-(\elmax-1)}^{\elmax-1}}}
\newcommand{\summptrunc}{\ensuremath{\sum_{\m\p=-(\elmax-1)}^{\elmax-1}}}
\newcommand{\qweight}{\ensuremath{q}}
\newcommand{\qweightdh}{\ensuremath{\qweight_{\rm DH}}}
\newcommand{\qweightmw}{\ensuremath{\qweight_{\rm MW}}}
\newcommand{\weighttrans}{\ensuremath{v}}

% CMB polarisation
\newcommand{\stokesi}{\ensuremath{I}}
\newcommand{\stokesq}{\ensuremath{Q}}
\newcommand{\stokesu}{\ensuremath{U}}
\newcommand{\stokesv}{\ensuremath{V}}
\newcommand{\emode}{\ensuremath{E}}
\newcommand{\bmode}{\ensuremath{B}}
\newcommand{\emodetilde}{\ensuremath{\tilde{E}}}
\newcommand{\bmodetilde}{\ensuremath{\tilde{B}}}
\newcommand{\qpmiu}{\ensuremath{\stokesq \pm \img \stokesu}}
\newcommand{\qpiu}{\ensuremath{\stokesq + \img \stokesu}}
\newcommand{\qmiu}{\ensuremath{\stokesq - \img \stokesu}}

% Numerics
\newcommand{\cswtfftterm}{\ensuremath{T}}
\newcommand{\grideula}{\ensuremath{\mathcal{E}_1}}
\newcommand{\grideulb}{\ensuremath{\mathcal{E}_2}}
\newcommand{\gridhpix}{\ensuremath{\mathcal{H}}}
\newcommand{\gridecp}{\ensuremath{\mathcal{C}}}
\newcommand{\ia}{\ensuremath{{n_\eula}}}
\newcommand{\ib}{\ensuremath{{n_\eulb}}}
\newcommand{\ig}{\ensuremath{{n_\eulc}}}
\newcommand{\ik}{\ensuremath{{k}}}
\newcommand{\ith}{\ensuremath{{n_\saa}}}
\newcommand{\iph}{\ensuremath{{n_\sab}}}
\newcommand{\ipix}{\ensuremath{{p}}}
\newcommand{\na}{\ensuremath{{N_\eula}}}
\newcommand{\nb}{\ensuremath{{N_\eulb}}}
\newcommand{\ngm}{\ensuremath{{N_\eulc}}}
\newcommand{\nth}{\ensuremath{{N_\saa}}}
\newcommand{\nph}{\ensuremath{{N_\sab}}}
\newcommand{\N}{\ensuremath{{N}}}
\newcommand{\Ngl}{\ensuremath{{N_{\rm GL}}}}
\newcommand{\Ndh}{\ensuremath{{N_{\rm DH}}}}
\newcommand{\Nhw}{\ensuremath{{N_{\rm HW}}}}
\newcommand{\Nmw}{\ensuremath{{N_{\rm MW}}}}
\newcommand{\Nm}{\ensuremath{{N_{\rm M}}}}
\newcommand{\Ns}{\ensuremath{{N_{\rm S}}}}
\newcommand{\nside}{\ensuremath{{N_{\rm{side}}}}}
\newcommand{\npix}{\ensuremath{{N_{\rm{pix}}}}}
\newcommand{\pixarea}{\ensuremath{{\Omega_{\rm{pix}}}}}
\newcommand{\nplane}{\ensuremath{{N_{\pind}}}}
\newcommand{\nsphere}{\ensuremath{{N_{\sind}}}}
\newcommand{\weight}{\ensuremath{w}}
\newcommand{\order}{\ensuremath{\mathcal{O}}}

% Skewness and kurtosis related variables
\newcommand{\mean}{\ensuremath{\mu}}
\newcommand{\skewness}{\ensuremath{\zeta}}
\newcommand{\kurtosis}{\ensuremath{\kappa}}
\newcommand{\neff}{\ensuremath{N_{\rm eff}}}
\newcommand{\num}{\ensuremath{N}}
\newcommand{\nstd}{\ensuremath{\num_\sigma}}
\newcommand{\ndev}{\ensuremath{\num_{\rm dev}}}
\newcommand{\nstat}{\ensuremath{\num_{\rm stat}}}
\newcommand{\conflevel}{\ensuremath{\rm SL}}
\newcommand{\cov}{\ensuremath{\mathbf{C}}}
\newcommand{\tstat}{\ensuremath{\tau}}

% Bianchi
\newcommand{\bx}{\ensuremath{x}}
\newcommand{\bhand}{\ensuremath{\kappa}}
\newcommand{\bh}{\ensuremath{h}}
\newcommand{\ze}{\ensuremath{\zrec}}
\newcommand{\bshear}{\ensuremath{\left(\frac{\sigma}{H}\right)_0}}
\newcommand{\bsheardim}{\ensuremath{\sigma}}
\newcommand{\bshearonetwo}{\ensuremath{\left(\frac{\sigma_{12}}{H}\right)_0}}
\newcommand{\bshearonethree}{\ensuremath{\left(\frac{\sigma_{13}}{H}\right)_0}}
\newcommand{\bvort}{\ensuremath{\left(\frac{\omega}{H}\right)_0}}
\newcommand{\bshearinline}{\ensuremath{(\sigma/H)_0}}
\newcommand{\bvortinline}{\ensuremath{(\omega/H)_0}}
\newcommand{\ba}{\ensuremath{A}}
\newcommand{\bb}{\ensuremath{B}}
\newcommand{\bi}[2]{\ensuremath{I^{#1}_{#2}}}
\newcommand{\thetaph}{\ensuremath{\theta_0}}
\newcommand{\phiph}{\ensuremath{\phi_0}}
\newcommand{\thetaob}{\ensuremath{\theta_{\rm ob}}}
\newcommand{\phiob}{\ensuremath{\phi_{\rm ob}}}
\newcommand{\thetaalm}{\ensuremath{\theta}}
\newcommand{\phialm}{\ensuremath{\phi}}
\newcommand{\bs}{\ensuremath{s}}
\newcommand{\bt}{\ensuremath{\tau}}
\newcommand{\bps}{\ensuremath{\psi}}
\newcommand{\bcone}{\ensuremath{C_1}}
\newcommand{\bctwo}{\ensuremath{C_2}}
\newcommand{\bcthree}{\ensuremath{C_3}}
\newcommand{\alm}{\ensuremath{a}}
\newcommand{\almi}{\ensuremath{\shc{\alm}{\el}{\m}}}
\newcommand{\almpi}{\ensuremath{\shc{\alm}{\el}{\m\p}}}
\newcommand{\almitilde}{\ensuremath{\shc{\tilde{\alm}}{\el}{\m}}}

% Bianchi template fitting
\newcommand{\amptmpl}{\ensuremath{\lambda}}
\newcommand{\prob}{\ensuremath{{\rm P}}}
\newcommand{\given}{\ensuremath{{\,\vert\,}}}
\newcommand{\evidence}{\ensuremath{{E}}}
\newcommand{\param}{\ensuremath{\Theta}}
\newcommand{\bparam}{\ensuremath{\Theta_{\rm B}}}
\newcommand{\cosmoparam}{\ensuremath{\Theta_{\rm C}}}
\newcommand{\anisoparam}{\ensuremath{\Theta_{\rm A}}}
\newcommand{\fitdata}{\ensuremath{\vect{d}}}
\newcommand{\fitmodel}{\ensuremath{\vect{m}}}
\newcommand{\fitmodelsel}{\ensuremath{M}}
\newcommand{\fitamp}{\ensuremath{\lambda}}
\newcommand{\fitampest}{\ensuremath{\widehat{\lambda}}}
\newcommand{\fittmpl}{\ensuremath{\vect{b}}}
\newcommand{\fitnoise}{\ensuremath{\vect{n}}}
\newcommand{\fitcov}{\ensuremath{\mathbf{M}}}
\newcommand{\fitchisqd}{\ensuremath{\chi^2}}
\newcommand{\fitdatasky}{\ensuremath{d}}
\newcommand{\fitdataalm}{\ensuremath{\shc{d}{\el}{\m}}}
\newcommand{\fittmplalm}{\ensuremath{\shc{b}{\el}{\m}}}
\newcommand{\fitdataalz}{\ensuremath{\shc{d}{\el}{0}}}
\newcommand{\fittmplalz}{\ensuremath{\shc{b}{\el}{0}}}
\newcommand{\mnoise}{\ensuremath{m}}

% ISW/Cross-correlation
\newcommand{\wcov}{\ensuremath{X_\wav}}
\newcommand{\wcovest}{\ensuremath{\widehat{X}_\wav}}
\newcommand{\wcovvect}{\ensuremath{\mathbf{X}_\wav}}
\newcommand{\wcovestvect}{\ensuremath{\widehat{\mathbf{X}}_\wav}}
\newcommand{\covweight}{\ensuremath{\nu}}
\newcommand{\covsubterm}{\ensuremath{G^\el}}
\newcommand{\clnt}{\ensuremath{C_\el^{\rm NT, obs}}}
\newcommand{\clnttheo}{\ensuremath{C_\el^{\rm NT}}}
\newcommand{\clnntheo}{\ensuremath{C_\el^{\rm NN}}}
\newcommand{\cltttheo}{\ensuremath{C_\el^{\rm TT}}}
\newcommand{\ndlab}{\ensuremath{{\rm N}}}
\newcommand{\tplab}{\ensuremath{{\rm T}}}
\newcommand{\dndz}{\ensuremath{\frac{\dx N}{\dx z}}}
\newcommand{\dgdz}{\ensuremath{\frac{\dx g}{\dx z}}}
\newcommand{\eqdec}{\ensuremath{\delta}}
\newcommand{\chisqd}{\ensuremath{\chi^2}}
\newcommand{\baypar}{\ensuremath{\Theta}}
\newcommand{\lhood}{\ensuremath{\mathcal{L}}}

% Optimal filters
\newcommand{\fil}{\ensuremath{\varphi}}
\newcommand{\filcoeff}{\ensuremath{w}}
\newcommand{\lagrnmult}{\ensuremath{\mu}}
\newcommand{\lagrn}{\ensuremath{\mathcal{L}}}
\newcommand{\filvara}{\ensuremath{a}}
\newcommand{\filvarb}{\ensuremath{b}}
\newcommand{\filvarc}{\ensuremath{c}}
\newcommand{\filvardenom}{\ensuremath{\Delta}}
\newcommand{\pnorm}{\ensuremath{p}}
\newcommand{\pnormsep}{\ensuremath{|}}
\newcommand{\pnormtext}{\ensuremath{\pnorm}-norm}
\newcommand{\pnormtextit}{\ensuremath{\rm{\pnorm}}-norm}
\newcommand{\scalepnorm}{\ensuremath{{\scale\pnormsep\pnorm}}}
\newcommand{\scalenautpnorm}{\ensuremath{{\scalenaut\pnormsep\pnorm}}}
\newcommand{\za}[2]{\ensuremath{\shc{a}{#1}{#2}}}
\newcommand{\zb}[2]{\ensuremath{\shc{b}{#1}{#2}}}
\newcommand{\zc}[2]{\ensuremath{\shc{c}{#1}{#2}}}
\newcommand{\zd}[2]{\ensuremath{\shc{d}{#1}{#2}}}
\newcommand{\zcsp}[2]{\ensuremath{\shcsp{c}{#1}{#2}}}
\newcommand{\zdsp}[2]{\ensuremath{\shcsp{d}{#1}{#2}}}
\newcommand{\detect}{\ensuremath{\Gamma}}
\newcommand{\source}{\ensuremath{s}}
\newcommand{\noise}{\ensuremath{n}}
\newcommand{\instnoise}{\ensuremath{r}}
\newcommand{\obsig}{\ensuremath{y}}
\newcommand{\conv}{\ensuremath{\star}}
\newcommand{\convdir}{\ensuremath{\circledast}}
\newcommand{\convaxisym}{\ensuremath{\odot}}
\newcommand{\amp}{\ensuremath{A}}
\newcommand{\tmpl}{\ensuremath{\tau}}
\newcommand{\noisecl}{\ensuremath{C}}
\newcommand{\amphat}{\ensuremath{\hat{A}}}
\newcommand{\eulshat}{\ensuremath{\hat{\alpha},\hat{\beta},\hat{\gamma}}}
\newcommand{\fconsta}{\ensuremath{A_{\el\pnorm}}}
\newcommand{\fconstb}{\ensuremath{B_{\el\m}}}

% Interferometry
\newcommand{\sao}{\ensuremath{\sa_0}}
\newcommand{\saog}{\ensuremath{\sao^{\glob}}}
\newcommand{\sag}{\ensuremath{\sa^{\glob}}}
\newcommand{\sal}{\ensuremath{\sa^{\loc}}}
\newcommand{\sae}{\ensuremath{\sa^{\erth}}}
\newcommand{\san}{\ensuremath{\sa^{\nght}}}
\newcommand{\saoe}{\ensuremath{\sao^{\erth}}}
\newcommand{\saon}{\ensuremath{\sao^{\nght}}}
\newcommand{\salcoordfree}{\ensuremath{\vect{\tau}}}
\newcommand{\vis}{\ensuremath{y}}
\newcommand{\visdis}{\ensuremath{\vect{y}}}
\newcommand{\beam}{\ensuremath{A}}
\newcommand{\beamp}{\ensuremath{A_{\pind}}}
\newcommand{\beams}{\ensuremath{A_{\sind}}}
\newcommand{\beamdis}{\ensuremath{\vect{\beam}}}
\newcommand{\imdis}{\ensuremath{\vect{\im}}}
\newcommand{\noisedis}{\ensuremath{\vect{\noise}}}
\newcommand{\imp}{\ensuremath{\im_{\pind}}}
\newcommand{\ims}{\ensuremath{\im_{\sind}}}
\newcommand{\imm}{\ensuremath{\im_{\mind}}}
\newcommand{\nim}{\ensuremath{n}}
\newcommand{\visvect}{\ensuremath{\vect{\vis}}}
\newcommand{\imvect}{\ensuremath{\vect{\im}}}
\newcommand{\imvectrecon}{\ensuremath{\vect{\im}^{\star}}}
\newcommand{\impvect}{\ensuremath{\vect{\imp}}}
\newcommand{\imsvect}{\ensuremath{\vect{\ims}}}
\newcommand{\immvect}{\ensuremath{\imvect_{m}}}
\newcommand{\impvectrecon}{\ensuremath{\vect{\imp}^{\star}}}
\newcommand{\imsvectrecon}{\ensuremath{\vect{\ims}^{\star}}}
\newcommand{\nimvect}{\ensuremath{\vect{\nim}}}
\newcommand{\bu}{\ensuremath{u}}
\newcommand{\bv}{\ensuremath{v}}
\newcommand{\bw}{\ensuremath{w}}
\newcommand{\bwd}{\ensuremath{w_{\rm d}}}
\newcommand{\buvect}{\ensuremath{\vect{\bu}}}
\newcommand{\buvectfull}{\ensuremath{\vect{b}}}
\newcommand{\bumax}{\ensuremath{\bu_{\rm max}}}
\newcommand{\bumaxfov}{\ensuremath{\bumax}}
\newcommand{\elmaxfov}{\ensuremath{\elmax}}
\newcommand{\lx}{\ensuremath{l}}
\newcommand{\mx}{\ensuremath{m}}
\newcommand{\nx}{\ensuremath{n}}
\newcommand{\lxvect}{\ensuremath{\vect{\lx}}}
\newcommand{\lxvectdis}{\ensuremath{\vect{\lx}_{\ispar}}}
\newcommand{\blinel}{\ensuremath{\vect{u}}}
\newcommand{\blineg}{\ensuremath{\vect{u}^\glob}}
\newcommand{\blinee}{\ensuremath{\vect{u}^\erth}}
\newcommand{\inten}{\ensuremath{I}}
\newcommand{\loc}{\ensuremath{{\rm l}}}
\newcommand{\glob}{\ensuremath{{\rm n}}}
\newcommand{\erth}{\ensuremath{{\rm e}}}
\newcommand{\nght}{\ensuremath{{\rm n}}}
\newcommand{\beaml}{\ensuremath{\beam^\loc}}
\newcommand{\beamg}{\ensuremath{\beam^\glob}}
\newcommand{\intenl}{\ensuremath{\inten^\loc}}
\newcommand{\inteng}{\ensuremath{\inten^\glob}}
\newcommand{\intenn}{\ensuremath{\inten^\nght}}
\newcommand{\beammodintenl}{\ensuremath{\mbox{$\bigl( \beaml \cdot \intenl \bigr)$}}}
\newcommand{\beamhorzmodintenl}{\ensuremath{\mbox{$\bigl( \beaml \cdot \horizonl \cdot \intenl \bigr)$}}}
\newcommand{\blinemax}{\ensuremath{\|\vect{u}\|_{\rm max}}}
\newcommand{\horizon}{\ensuremath{H}}
\newcommand{\horizone}{\ensuremath{\horizon^\erth}}
\newcommand{\horizonl}{\ensuremath{\horizon^\loc}}
\newcommand{\chirp}{\ensuremath{C}}
\newcommand{\chirpfull}{\ensuremath{\chirp(\| \lxvect \|_2)}}
\newcommand{\chirpfulla}{\ensuremath{\chirp_{0}(\| \lxvect \|_2)}}
\newcommand{\chirpfullb}{\ensuremath{\chirp_{1}(\| \lxvect \|_2)}}
\newcommand{\chirpdis}{\ensuremath{\vect{\chirp}}}

\newcommand{\chirpfullbl}{\ensuremath{\bumax^{\chirp^{(\bw)}}}}
\newcommand{\chirpfullbbl}{\ensuremath{\bumax^{\chirp^{(\bw)}_1}}}

\newcommand{\expcoeff}{\ensuremath{E}}
\newcommand{\objsize}{\ensuremath{\sigma_{\rm S}}}
\newcommand{\plevel}{\ensuremath{p}}
\newcommand{\pfov}{\ensuremath{L}}

% Compressed sensing
\newcommand{\nmeas}{\ensuremath{M}}
\newcommand{\ndim}{\ensuremath{N}}
\newcommand{\sparsity}{\ensuremath{K}}
\newcommand{\ssmat}{\ensuremath{\Theta}}
\newcommand{\ispar}{\ensuremath{i}}
\newcommand{\isens}{\ensuremath{r}}
\newcommand{\sparatom}{\ensuremath{\vect{\psi}_\ispar}}
\newcommand{\sensatom}{\ensuremath{\vect{\phi}_\isens}}
\newcommand{\sparmat}{\ensuremath{\Psi}}
\newcommand{\sensmat}{\ensuremath{\Phi}}
\newcommand{\sensmatm}{\ensuremath{\Phi_{m}}}
\newcommand{\sensmatp}{\ensuremath{\Phi_{\pind}}}
\newcommand{\sensmats}{\ensuremath{\Phi_{\sind}}}
\newcommand{\coherence}{\ensuremath{\mu}}
\newcommand{\ccoherence}{\ensuremath{\nu}}
\newcommand{\coherences}{\ensuremath{\coherence _{\rm s}}}
\newcommand{\ccoherences}{\ensuremath{\ccoherence _{\rm s}}}
\newcommand{\coherencep}{\ensuremath{\coherence _{\rm p}}}
\newcommand{\ccoherencep}{\ensuremath{\ccoherence _{\rm p}}}

% Compressed sensing for interferometry
\newcommand{\opmask}{\ensuremath{\mathbfss{M}}}
\newcommand{\opbeam}{\ensuremath{\mathbfss{A}}}
\newcommand{\opfourier}{\ensuremath{\mathbfss{F}}}
\newcommand{\opwhite}{\ensuremath{\mathbfss{W}}}
\newcommand{\opchirp}{\ensuremath{\mathbfss{C}}}
\newcommand{\opgrid}{\ensuremath{\mathbfss{G}}}
\newcommand{\opproj}{\ensuremath{\mathbfss{P}}}

% Sparse signal reconstruction on the sphere
\newcommand{\opshtinv}{\ensuremath{\Lambda}}
\newcommand{\opshtfwd}{\ensuremath{\Gamma}}
\newcommand{\opshtcse}{\ensuremath{\Pi}}

% Bubble collisions
\newcommand{\thetacrit}{\ensuremath{\saa_{\rm crit}}}
\newcommand{\zo}{\ensuremath{z_{0}}}
\newcommand{\zcrit}{\ensuremath{z_{\rm crit}}}
\newcommand{\saao}{\ensuremath{\saa_{0}}}
\newcommand{\sabo}{\ensuremath{\sab_{0}}}
\newcommand{\saso}{\ensuremath{\saao, \sabo}}

% Cosmic strings
\newcommand{\gmu}{\ensuremath{G \mu}}
\newcommand{\fstring}{\ensuremath{s}}
\newcommand{\fcmb}{\ensuremath{c}}
\newcommand{\fdata}{\ensuremath{d}}
\newcommand{\fnoise}{\ensuremath{n}}
\newcommand{\fcmbnoise}{\ensuremath{{g}}}
\newcommand{\fstringrecov}{\ensuremath{\overline{\fstring}}}
\newcommand{\wcoeffstring}{\ensuremath{\wcoeff^\fstring}}
\newcommand{\wcoeffcmb}{\ensuremath{\wcoeff^\fcmb}}
\newcommand{\wcoeffdata}{\ensuremath{\wcoeff^\fdata}}
\newcommand{\wcoeffnoise}{\ensuremath{\wcoeff^\fnoise}}
\newcommand{\wcoeffcmbnoise}{\ensuremath{\wcoeff^{\fcmbnoise}}}
\newcommand{\wcoeffstringp}{\ensuremath{\wcoeffstring_{\wscaleposn}}}
\newcommand{\wcoeffcmbp}{\ensuremath{\wcoeffcmb_{\wscaleposn}}}
\newcommand{\wcoeffdatap}{\ensuremath{\wcoeffdata_{\wscaleposn}}}
\newcommand{\wcoeffnoisep}{\ensuremath{\wcoeffnoise_{\wscaleposn}}}
\newcommand{\wcoeffcmbnoisep}{\ensuremath{\wcoeffcmbnoise_{\wscaleposn}}}
\newcommand{\wcoeffstringprecov}{\ensuremath{\overline{\wcoeff}^\fstring_{\wscaleposn}}}
\newcommand{\wcoeffstringrecov}{\ensuremath{\overline{\wcoeff}^\fstring}}
\newcommand{\wvarcmb}{\ensuremath{(\sigma^\fcmb_\wscale)^2}}
\newcommand{\wstdcmb}{\ensuremath{\sigma^\fcmb_\wscale}}
\newcommand{\wvarstring}{(\sigma^\fstring_\wscale)^2}
\newcommand{\wstdstring}{\ensuremath{\sigma^\fstring_\wscale}}
\newcommand{\wkurstring}{\ensuremath{\kurtosis^\fstring_\wscale}}
\newcommand{\ggdshape}{\ensuremath{\changed{\xi}}}
\newcommand{\ggdscale}{\ensuremath{\changed{\zeta}}}
\newcommand{\ggdshapew}{\ensuremath{{\ggdshape_\wscale}}}
\newcommand{\ggdscalew}{\ensuremath{{\ggdscale_\wscale}}}
\newcommand{\modelstring}{\ensuremath{{{\rm M}^{\rm \fstring}}}}
\newcommand{\modelcmb}{\ensuremath{{{\rm M}^{\rm \fcmb}}}}

% Weak lensing
\newcommand{\wlconv}{\ensuremath{\prescript{}{0}{\kappa}}}
\newcommand{\wlshear}{\ensuremath{\prescript{}{2}{\gamma}}}
\newcommand{\wlshearone}{\ensuremath{\gamma_1}}
\newcommand{\wlsheartwo}{\ensuremath{\gamma_2}}
\newcommand{\wlpot}{\ensuremath{\prescript{}{0}{\phi}}}

\newcommand{\wlconvshc}{\ensuremath{\sshc{\hat{\kappa}}{\el}{\m}{0}}}
\newcommand{\wlshearshc}{\ensuremath{\sshc{\hat{\gamma}}{\el}{\m}{2}}}
\newcommand{\wlpotshc}{\ensuremath{\sshc{\hat{\phi}}{\el}{\m}{0}}}

\newcommand{\wlconvf}{\ensuremath{\prescript{}{0}{\hat{\kappa}}}}
\newcommand{\wlshearf}{\ensuremath{{\hat{\gamma}_1}}}
\newcommand{\wlshearonef}{\ensuremath{{\hat{\gamma}_1}}}
\newcommand{\wlsheartwof}{\ensuremath{{\hat{\gamma}_2}}}
\newcommand{\wlpotf}{\ensuremath{\prescript{}{0}{\hat{\phi}}}}

\newcommand{\wlconvfk}{\ensuremath{\prescript{}{0}{\hat{\kappa}(\wlkx,\wlky)}}}
\newcommand{\wlshearfk}{\ensuremath{\prescript{}{2}{\hat{\gamma}(\wlkx,\wlky)}}}
\newcommand{\wlshearonefk}{\ensuremath{{\hat{\gamma}_1(\wlkx,\wlky)}}}
\newcommand{\wlsheartwofk}{\ensuremath{{\hat{\gamma}_2(\wlkx,\wlky)}}}
\newcommand{\wlpotfk}{\ensuremath{\prescript{}{0}{\hat{\phi}(\wlkx,\wlky)}}}

\newcommand{\wlkx}{\ensuremath{k_x}}
\newcommand{\wlky}{\ensuremath{k_y}}

\newcommand{\wlfactor}{\ensuremath{\mathcal{D}}}
\newcommand{\wlfactorplane}{\ensuremath{\mathcal{E}}}
\newcommand{\wlkernel}{\ensuremath{\prescript{}{2}{\mathcal{K}}}}

\renewcommand{\prob}{\ensuremath{\text{P}}}
\renewcommand{\evidence}{\ensuremath{{z}}}
\newcommand{\evidenceinv}{\ensuremath{\rho}}
\newcommand{\data}{\ensuremath{{y}}}
\newcommand{\datum}{\ensuremath{y}}
\renewcommand{\exp}{\ensuremath{\text{exp}}}

\def\maxdim{10000}

\title[Learnt harmonic mean estimator]{Machine learning assisted Bayesian model comparison: learnt harmonic mean estimator  %\thanks{Grants or other notes
}

\author*[1,2]{\fnm{Jason}~D.~\sur{McEwen}}\email{jason.mcewen@ucl.ac.uk}

\author[1]{\fnm{\mbox{Christopher}}~G.~R.~\sur{Wallis}} %\email{iiauthor@gmail.com}

\author[1]{\fnm{Matthew}~A.~\sur{Price}}%\email{iiiauthor@gmail.com}

\author[1]{\fnm{Alessio}~\sur{Spurio Mancini}}%\email{iiiauthor@gmail.com}

\affil[1]{\orgdiv{Mullard Space Science Laboratory (MSSL)}, \orgname{University College London (UCL)}, \orgaddress{\city{Dorking}, \postcode{RH5 6NT}, \country{UK}}}

\affil[2]{\orgname{Alan Turing Institute}, \orgaddress{\city{London}, \postcode{NW1 2DB}, \country{UK}}}

% \affil[3]{\orgdiv{Department}, \orgname{Organization}, \orgaddress{\street{Street}, \city{City}, \postcode{610101}, \state{State}, \country{Country}}}

% \begin{abstract}
\abstract{We resurrect the infamous harmonic mean estimator for computing the marginal likelihood (Bayesian evidence) and solve its problematic large variance.  The marginal likelihood is a key component of Bayesian model selection to evaluate model posterior probabilities;  however, its computation is challenging.  The original harmonic mean estimator, first proposed by Newton and Raftery in 1994, involves computing the harmonic mean of the likelihood given samples from the posterior.  It was immediately realised that the original estimator can fail catastrophically since its variance can become very large (possibly not finite).  A number of variants of the harmonic mean estimator have been proposed to address this issue although none have proven fully satisfactory. We present the \emph{learnt harmonic mean estimator}, a variant of the original estimator that solves its large variance problem.  This is achieved by interpreting the harmonic mean estimator as importance sampling and introducing a new target distribution.  The new target distribution is learned to approximate the optimal but inaccessible target, while minimising the variance of the resulting estimator.  Since the estimator requires samples of the posterior only, it is agnostic to the sampling strategy used.  We validate the estimator on a variety of numerical experiments, including a number of pathological examples where the original harmonic mean estimator fails catastrophically.  We also consider a cosmological application, where our approach leads to $\sim$ 3 to 6 times more samples than current state-of-the-art techniques in 1/3 of the time.  In all cases our learnt harmonic mean estimator is shown to be highly accurate. The estimator is computationally scalable and can be applied to problems of dimension $\order(10^3)$ and beyond. Code implementing the learnt harmonic mean estimator is made publicly available.}
%
% \keywords{model selection \and Bayesian statistics \and marginal likelihood \and Bayesian evidence \and machine learning.}
% \end{abstract}

\maketitle

%===============================================================================
\section{Introduction}
%===============================================================================

Model selection is a critical task in order to ascertain an appropriate statistical model to describe observational data.  In the Bayesian formalism, model selection requires computing the \textit{marginal likelihood}, the average likelihood of a model over its prior probability space, given observational data.  The marginal likelihood (also called the \textit{Bayesian evidence}) may then be used to compute model posterior odds and assign relative probabilities to different models. Computing the marginal likelihood is therefore a key ingredient in Bayesian inference.  However, computing the marginal likelihood in practice requires the evaluation of a high-dimensional integral, which is computationally challenging.

The Bayesian formalism is one of the most common approaches to statistical inference.  Consider the estimation of unknown parameters $\theta \in \Theta$ (typically \mbox{$\Theta = \mathbb{R}^d$}) from observed data $\data$ (typically $\data \in \mathbb{R}^n$), under a statistical model $M$ relating the data to the parameters.  Given the data $\data$ and model $M$, inferences of the parameters $\theta$ are based on their posterior distribution through Bayes' theorem by
\begin{equation}
  \text{P}(\theta \given \data, M)
  = \frac{\text{P}(\data \given \theta, M) \text{P}(\theta \given M)}{\text{P}(\data \given M)}
  = \frac{\mathcal{L}(\theta) \pi(\theta)}{z}
  \spcend ,
\end{equation}
where for model $M$ the likelihood $\text{P}(\data \given \theta, M)$ specifies the probability of the data given the parameters and $\text{P}(\theta \given M)$ encodes prior information about the parameters. The denominator $\text{P}(\data \given M)$, termed the \textit{marginal likelihood} or \textit{Bayesian evidence}, measures the probability of the observed data under model $M$.  For notational brevity we denote the likelihood by $\mathcal{L}(\theta)$, the prior by $\pi(\theta)$ and the marginal likelihood by $z$.  We drop the explicit dependence on the model $M$, except where explicitly required.  For parameter inference the marginal likelihood can be ignored (since it simply normalises the posterior) and the shape of the posterior can be explored using Markov chain Monte Carlo (MCMC) sampling techniques, e.g.\ Metropolis-Hastings sampling \citep{metropolis:1953,hastings:1970}.

For Bayesian model selection it is necessary to compute the marginal likelihood given by
\begin{equation}
  \label{eqn:evidence}
  z =
  \text{P}(\data \given M)
  = \int \,\text{d} \theta \:
  \text{P}(\data \given \theta, M) \text{P}(\theta \given M)
  = \int \,\text{d} \theta \:
  \mathcal{L}(\theta) \pi(\theta)
  \spcend .
\end{equation}
The marginal likelihood is of critical importance for Bayesian model selection since it is required to compute the posterior probabilities of models.  Noting Bayes' theorem, the relative posterior probability of competing models $M_1$ and $M_2$ is given by
\begin{equation}
  \frac{\text{P}(M_1 \given \data)}{\text{P}(M_2 \given \data)}
  = \frac{\text{P}(\data \given M_1)}{\text{P}(\data \given M_2)}
  \frac{\text{P}(M_1)}{\text{P}(M_2)}
  \spcend .
\end{equation}
In the absence of prior information regarding model preferences it is reasonable to take the ratio of model prior probabilities $\text{P}(M_1) / \text{P}(M_2)$ to be unity.  In this case, the relative model posterior probability is given by the ratio of marginal likelihoods for the two competing models, which is also called the \textit{Bayes factor}.  In either case, computing marginal likelihoods is a critical component to evaluating model posterior odds, which can then be used to select the preferred model.

It is clear from \eqn{\ref{eqn:evidence}} that evaluation of the marginal likelihood requires computation of an integral with dimension $d$ given by the number of parameters of interest, which is typically {high-dimen\-sional}.
In principle, the marginal likelihood could be computed simply by Monte Carlo integration of the likelihood, given samples from the prior.  While this estimator converges asymptotically to the true marginal likelihood as the number of Monte Carlo samples increases, in practice the accuracy of the estimator depends critically on its variance.  Since in practice the prior is typically more diffuse than the likelihood this approach is inefficient, particularly in high and even moderate dimensional settings \citep{clyde:2007}.  Consequently, this simple estimator is usually not effective in practice \citep[see, e.g.,][]{cai:proximal_nested_sampling}.

A variety of alternative methods have been proposed to compute the marginal likelihood.  For excellent reviews see \citet{clyde:2007} and \citet{friel:2012}.
The Savage-Dickey density ratio can be used for nested models \citep{trotta:2007}.  For more general models, Laplace's method is a widely used approach \citep{tierney:1986}, which relies on the assumption that the posterior distribution can be adequately approximated by a Gaussian distribution.  This assumption often does not hold and so marginal likelihood estimates computed by Laplace's method may be inaccurate.
Thermodynamic integration (e.g.\ \citealt{ruanaidh:1996}), which is based on MCMC techniques, is a well-known, general approach for computing the marginal likelihood that has been applied successfully for low-dimensional problems \citep[e.g.][]{marshall:2003}; however, it does require careful tuning.
Annealed importance sampling \citep{neal:2001}, which approximates the target distribution using a tempering mechanism to adaptively define an importance sampling function, is another.  Chib's method \citep{chib:1995,chib:2001} is based on the outputs of a Gibbs or Metropolis-Hasting (MH) sampler \citep{metropolis:1953,hastings:1970}, which poses some restrictions.
Nested sampling \citep{skilling:2006} was designed specifically with the computation of the marginal likelihood in mind, reparameterising the marginal likelihood into a one-dimensional integral of the likelihood with respected to the enclosed prior volume.  The computational difficulty of nested sampling approaches is shifted to sampling of the prior distribution subject to a hard constraint defined by likelihood level-sets.  Numerous nested sampling strategies have been proposed based on MCMC sampling \citep{skilling:2006}, ellipsoidal rejection sampling \citep{feroz:multinest1,feroz:multinest2}, slice sampling \citep{handley:2015}, diffusive sampling \citep{brewer:2011} and proximal sampling \citep{cai:proximal_nested_sampling}.
In all of the above approaches, the sampling strategy is tightly coupled with the technique used to estimate the marginal likelihood.  Furthermore, while nested sampling approaches have scaled to high-dimensional settings, notably proximal nested sampling to dimensions $10^6$ and beyond \citep{cai:proximal_nested_sampling}, most techniques are limited to low-dimensional settings.

Ideally, the computation of the marginal likelihood would be agnostic to the sampling strategy.  If the marginal likelihood estimator required samples from the posterior only, it could indeed then be decoupled from sampling.  In this case, the most effective sampler for the problem at hand could be considered and the posterior samples recycled to estimate the marginal likelihood.  While some techniques to compute the marginal likelihood from posterior samples have been proposed, they are generally not robust and limited to very low dimensions.  The harmonic mean estimator \citep{newton:1994} involves computing the harmonic mean of the likelihood given samples of the posterior generated by any MCMC technique.  However, it was immediately realised that the original estimator can fail catastrophically since its variance can become very large and may not be finite (a thorough review and inspection of the harmonic mean estimator and variants is presented in \sectn{\ref{sec:review}}).  In \citet{heavens:2017} an approach based on $k$th nearest-neighbour distances is proposed to compute the marginal likelihood from posterior samples, although the technique is limited to low-dimensional settings.

In this article we present the \textit{learnt harmonic mean estimator}, a variant of the original harmonic mean estimator that solves its large variance problem.  This is achieved by interpreting the harmonic mean estimator as importance sampling and introducing a new target distribution.  The new target distribution is learned to approximate the optimal but inaccessible target, while minimising the variance of the resulting estimator.  The estimator requires samples of the posterior only and hence is agnostic to the strategy used to generate posterior samples.  Posterior samples are split to first learn the target distribution and then to second infer the marginal likelihood using the learnt target.  The resulting estimator is evaluated on a variety of numerical experiments, including a number of pathological examples where the original harmonic mean estimator has been shown to fail catastrophically.  In all cases our learnt harmonic mean estimator is shown to be robust and highly accurate.

The remainder of this article is structured as follows.  In \sectn{\ref{sec:review}} we review the harmonic mean estimator, its problematic source of large variance, and variants that have been introduced in an attempt to mitigate this issue.  We present our learnt harmonic mean estimator in \sectn{\ref{sec:learnt_harmonic_mean}}.  In \sectn{\ref{sec:experiments}} we apply our estimator to numerous benchmark problems where ground truth marginal likelihood values are accessible, demonstrating in all cases that it is highly accurate.  Particular attention has been paid to the design and implementation of the software code implementing our learnt harmonic mean estimator so that it can be easily applied by others to their problems of interest.  We demonstrate the ease of use of the code in \sectn{\ref{sec:code}}.  Concluding remarks are made in \sectn{\ref{sec:conclusions}}.

%===============================================================================
\section{Review of harmonic mean estimators}
\label{sec:review}
%===============================================================================

Harmonic mean estimators have been the focus of considerable discussion since first proposed by \citet{newton:1994}.  While the harmonic mean estimator is asymptotically consistent \citep{newton:1994}, it was immediately realised that the original estimator can fail catastrophically \citep{neal:1994} since its variance can become very large and may not be finite. A number of variants of the original estimator have been proposed to address its failings \citep[\eg][]{raftery:2006, robert:2009, lenk:2009, vanhaasteren:2014}, although harmonic mean estimators have generally been considered to be ineffective \citep{clyde:2007, friel:2012}.  We review the original harmonic mean estimator and discuss why it is problematic.  We then review variants of the original estimator and how they attempt to address this failing, which motivates the \emph{learnt harmonic mean estimator} that we present in \sectn{\ref{sec:learnt_harmonic_mean}}.

%-------------------------------------------------------------------------------
\subsection{Original harmonic mean estimator}
%-------------------------------------------------------------------------------

The harmonic mean estimator was first proposed by \citet{newton:1994}, who showed that the marginal likelihood $z$ can be estimated from the harmonic mean of the likelihood, given posterior samples.  This follows by considering the expectation of the reciprocal of the likelihood with respect to the posterior distribution:
\begin{align}
  \rho
   & =
  \mathbb{E}_{\text{P}(\theta \given \data)} \biggl[
    \frac{1}{\mathcal{L}(\theta)}
  \biggr]                                   \\
   & = \int \,\text{d} \theta
  \frac{1}{\mathcal{L}(\theta)}
  \text{P}(\theta \given \data)             \\
   & = \int \,\text{d} \theta
  \frac{1}{\mathcal{L}(\theta)}
  \frac{\mathcal{L}(\theta) \pi(\theta)}{z} \\
   & = \frac{1}{z}
  \spcend ,
\end{align}
where the final line follows since the prior $\pi(\theta)$ is a normalised probability distribution.  This relationship between the marginal likelihood and the harmonic mean motivates the \emph{original harmonic mean estimator}:
\begin{equation}
  \hat{\rho} = \frac{1}{N} \sum_{i=1}^{N} \frac{1}{\mathcal{L}(\theta_i)} \spcend ,
  \quad
  \theta_i \sim \text{P}(\theta \given \data)
  \spcend ,
\end{equation}
where $N$ specifies the number of samples $\theta_i$ drawn from the posterior, and from which the marginal likelihood may naively be estimated by $\hat{z} = 1 / \hat{\rho}$. For now we simply consider the estimation of the reciprocal of the marginal likelihood $\hat{\rho}$ (we discuss estimation of the marginal likelihood itself and Bayes factors in more detail in \sectn{\ref{sec:learnt_harmonic_mean:bayes_factors}}).

As immediately realised by \citet{neal:1994}, this estimator can fail catastrophically since its variance can become very large and may not be finite.  Review articles that consider a variety of methods to estimate the marginal likelihood have also found that the harmonic mean estimator is not robust and can be highly inaccurate \citep{clyde:2007, friel:2012}.  To understand why the estimator can lead to extremely large variance we consider an importance sampling interpretation of the harmonic mean estimator.

%-------------------------------------------------------------------------------
\subsubsection{Importance sampling interpretation}

The harmonic mean estimator can be interpreted as importance sampling.  Consider the reciprocal marginal likelihood, which may be expressed in terms of the prior and posterior by
\begin{align}
  \rho
   & = \int \,\text{d} \theta \:
  \frac{1}{\mathcal{L}(\theta)} \: \text{P}(\theta \given \data) \\
   & = \int \,\text{d} \theta \:
  \frac{1}{z} \:
  \frac{\pi(\theta)}{\text{P}(\theta \given \data)} \:
  \text{P}(\theta \given \data)
  \spcend .
\end{align}
It is clear the estimator has an importance sampling interpretation where the importance sampling target distribution is the prior $\pi(\theta)$, while the sampling density is the posterior $\text{P}(\theta \given \data)$, in contrast to typical importance sampling scenarios.

For importance sampling to be effective, one requires the sampling density to have fatter tails than the target distribution, i.e.\ to have greater probability mass in the tails of the distribution.  Typically the prior has fatter tails than the posterior since the posterior updates our initial understanding of the underlying parameters $\theta$ that are encoded in the prior, in the presence of new data $\data$.  For the harmonic mean estimator the importance sampling density (the posterior) typically does \emph{not} have fatter tails than the target (the prior) and so importance sampling is not effective.  This explains why the original harmonic mean estimator can be problematic. A number of variants of the original harmonic mean estimator have been introduced in an attempt to address this issue.

%-------------------------------------------------------------------------------
\subsection{Adjusted harmonic mean estimator}
%-------------------------------------------------------------------------------

\citet{lenk:2009} show that while the original harmonic mean estimator is consistent, in practice it exhibits simulation pseudo-bias.  Simulation pseudo-bias arises since the posterior simulation support is a subset of the prior support.  Consequently, the prior is not sufficiently captured, which often results in an over-estimate of the marginal likelihood.

An \emph{adjusted harmonic mean estimator} is introduced by \citet{lenk:2009} to correct for simulation pseudo-bias:
\begin{equation}
  \hat{\rho} = \frac{1}{\text{P}(\Lambda)} \frac{1}{N} \sum_{i=1}^{N} \frac{1}{\mathcal{L}(\theta_i)} \spcend ,
  \quad
  \theta_i \sim \text{P}(\theta \given \data)
  \spcend ,
\end{equation}
where $\text{P}(\Lambda)$ is a pseudo-bias adjustment factor given by the prior probability of the posterior simulation support $\Lambda \subset \Theta$.  Numerical methods to estimate $\text{P}(\Lambda)$ are proposed, however, estimating the adjustment factor accurately is numerically challenging, particularly in high dimensions.  Furthermore, while this adjusted estimator can mitigate simulation pseudo-bias it does not eliminate it \citep{pajor:2017}. Alternative approaches seek to eliminate the bias altogether.

%-------------------------------------------------------------------------------
\subsection{Stabilised harmonic mean estimator}
%-------------------------------------------------------------------------------

\citet{raftery:2006} propose an alternative approach, a \emph{stabilised harmonic mean estimator}, by introducing a variance stabilisation strategy that reduces the size of the parameter space.  While this strategy can be applied to a variety of common hierarchical models it is not applicable in general, limiting its use.

%-------------------------------------------------------------------------------
\subsection{Re-targeted harmonic mean estimator}
\label{sec:review:re-targeted}
%-------------------------------------------------------------------------------

The original harmonic mean estimator was revised by \citet{gelfand:1994} by introducing an arbitrary density $\varphi(\theta)$ to relate the reciprocal of the marginal likelihood to the likelihood through the following expectation:
\begin{align}
  \rho
   & =
  \mathbb{E}_{\text{P}(\theta \given \data)} \biggl[
    \frac{\varphi(\theta)}{\mathcal{L}(\theta) \pi(\theta)}
  \biggr]                                   \\
   & = \int \,\text{d} \theta
  \frac{\varphi(\theta)}{\mathcal{L}(\theta) \pi(\theta)}
  \text{P}(\theta \given \data)             \\
   & = \int \,\text{d} \theta
  \frac{\varphi(\theta)}{\mathcal{L}(\theta) \pi(\theta)}
  \frac{\mathcal{L}(\theta) \pi(\theta)}{z} \\
   & = \frac{1}{z}
  \spcend ,
\end{align}
where the final line follows since the density $\varphi(\theta)$ must be normalised.
The above expression motivates the estimator:
\begin{equation}
  \label{eqn:harmonic_mean_retargeted}
  \hat{\rho} =
  \frac{1}{N} \sum_{i=1}^N
  \frac{\varphi(\theta_i)}{\mathcal{L}(\theta_i) \pi(\theta_i)} \spcend ,
  \quad
  \theta_i \sim \text{P}(\theta \given \data)
  \spcend .
\end{equation}
The normalised density $\varphi(\theta)$ can be interpreted as an alternative importance sampling target distribution, as we will see, hence we refer to this approach as the \emph{re-targeted harmonic mean estimator}.  Note that the original harmonic mean estimator is recovered for the target distribution $\varphi(\theta) = \pi(\theta)$.

%-------------------------------------------------------------------------------
\subsubsection{Importance sampling interpretation}
\label{sec:review:re-targeted:importance}

With the introduction of the distribution $\varphi(\theta)$, the importance sampling interpretation of the harmonic mean estimator reads
\begin{align}
  \rho
   & = \int \,\text{d} \theta \:
  \frac{\varphi(\theta)}{\mathcal{L}(\theta) \pi(\theta)} \:
  \text{P}(\theta \given \data)  \\
   & = \int \,\text{d} \theta \:
  \frac{1}{z} \:
  \frac{\varphi(\theta)}{\text{P}(\theta \given \data)} \:
  \text{P}(\theta \given \data)
  \spcend .
\end{align}
It is clear that the distribution $\varphi(\theta)$ now plays the role of the importance sampling target distribution.
One is free to choose $\varphi(\theta)$, with the only constraint being that it is a normalised distribution.  It is therefore possible to select the target distribution $\varphi(\theta)$ such that it has narrower tails than the posterior, which we recall plays the role of the importance sampling density, thereby avoiding the problematic scenario of the original harmonic mean estimator. We therefore refer to $\varphi(\theta)$ as the \textit{target distribution} of the harmonic mean estimator.

The question of how to develop an effective strategy to select $\varphi(\theta)$ for a given problem remains, which is particularly difficult in high-dimensional settings \citep{chib:1995}.  \citet{gelfand:1994} initially suggest using a multivariate Gaussian, although this approach is typically not effective since the tails of the distribution are generally not sufficiently narrow \citep{chib:1995, clyde:2007}.

%-------------------------------------------------------------------------------
\subsubsection{Truncated harmonic mean estimator}

A common strategy to select the target distribution is to set it to a normalised indicator function that is supported on a region $\Omega$ of high posterior mass (so that the target has narrower tails than the posterior):
\begin{equation}
  \varphi(\theta) = \frac{1}{V_\Omega} \: I_\Omega(\theta)
  \spcend ,
\end{equation}
where $V_\Omega$ represents the volume encapsulated in $\Omega$ and the indicator function $I_\Omega(\theta) = 1$ if $\theta \in \Omega$ and zero otherwise.  Since the indicator function effectively truncates the region of parameter space considered we refer to this approach as the \emph{truncated harmonic mean estimator}.

\citet{robert:2009} propose a target distribution that corresponds to an indicator function with support $\Omega$ determined from the convex hull of Monte Carlo samples within an $\alpha$\% highest posterior density (HPD) region.  In practice they consider an ellipsoidal region defined by HPD samples, for which the volume can be computed analytically.  \citet{vanhaasteren:2014} take a similar approach and and consider indicator functions defined over ellipsoidal regions.

While such approaches can be effective, in general the truncated harmonic mean estimator can be inaccurate and inefficient since each sample is either used, with a uniform target density weight, or discarded.  In scenarios that exhibit thin parameter degeneracies such approaches either capture large regions of low posterior mass, which is problematic (for reasons discussed above in \sectn{\ref{sec:review:re-targeted:importance}}), or can suffer prohibitive inefficiencies as the support of the target distribution $\Omega$ can be a very small region of the full parameter space $\Theta$ (resulting in very few samples being retained in the marginal likelihood computation).

The selection of appropriate target densities $\varphi(\theta)$ for general problems remains an open question that is known to be difficult, particularly in high dimensions \citep{chib:1995}.  One may gain insight into effective strategies to design the target density by considering the optimal target distribution.

%-------------------------------------------------------------------------------
\subsubsection{Optimal importance sampling target}

Consider the importance sampling target distribution given by the (normalised) posterior itself:
\begin{equation}
  \varphi^\text{optimal}(\theta) = \frac{\mathcal{L}(\theta) \pi(\theta)}{z}
  \spcend .
\end{equation}
This estimator is optimal in the sense of having zero variance, which is clearly apparent by substituting the target density into the re-targeted harmonic mean estimator of \eqn{\ref{eqn:harmonic_mean_retargeted}}. Each term contributing to the summation is simply $1/z$, hence the estimator $\hat{\rho}$ is unbiased, with zero variance.

Recall that the target density must be normalised.  Hence, the optimal estimator given by the normalised posterior is not accessible in practice since it requires the marginal likelihood -- the very term we are attempting to estimate -- to be known.  While the optimal estimator therefore cannot be used in practice, it can nevertheless be used to inform the construction of other estimators based on alternative importance sampling target distributions.

%===============================================================================
\section{Learnt harmonic mean estimator}
\label{sec:learnt_harmonic_mean}
%===============================================================================

It is well-known that the original harmonic mean estimator can fail catastrophically since the variance of the estimator may be become very large, as discussed in detail in \sectn{\ref{sec:review}}.  As also discussed in \sectn{\ref{sec:review}}, however, this issue can be resolved by introducing an alternative (normalised) target distribution $\varphi(\theta)$ \citep{gelfand:1994}, yielding what we term here the \textit{re-targeted harmonic mean estimator}.  From the importance sampling interpretation of the harmonic mean estimator, the re-targeted estimator follows by replacing the importance sampling target of the prior $\pi(\theta)$ with the target $\varphi(\theta)$, where the posterior $\prob(\theta \given \data)$ plays the role of the importance sampling density.

It remains to select a suitable target distribution $\varphi(\theta)$.  On one hand, to ensure the variance of the resulting estimator is well-behaved, the target distribution should have narrower tails that the importance sampling density, i.e.\ the target $\varphi(\theta)$ should have narrower tails than the posterior $\prob(\theta \given \data)$ (as discussed in \sectn{\ref{sec:review}}).  On the other hand, to ensure the resulting estimator is efficient and makes use of as many samples from the posterior as possible, the target distribution should not be too narrow.  The optimal target distribution is the normalised posterior distribution since in this case the variance of the resulting estimator is zero (\sectn{\ref{sec:review}}).  However, the normalised posterior is not accessible since it requires knowledge of the marginal likelihood, which is precisely the term we are attempting to compute.

We propose learning the target distribution $\varphi(\theta)$ from samples of the posterior.  Samples from the posterior can be split into training and evaluation (cf.\ test) sets.  Machine learning (ML) techniques can then be applied to learn an approximate model of the normalised posterior from the training samples, with the constraint that the tails of the learnt target are narrower than the posterior, i.e.\
\begin{equation}
  \varphi(\theta) \stackrel{\text{ML}}{\simeq} \varphi^\text{optimal}(\theta) = \frac{\mathcal{L}(\theta) \pi(\theta)}{z}
  \spcend .
\end{equation}
We term this approach the \textit{learnt harmonic mean estimator}.

We are interested not only in an estimator for the marginal likelihood but also in an estimate of the variance of this estimator, and its variance.  Such additional estimators are useful in their own right and can also provide valuable sanity checks that the resulting marginal likelihood estimator is well-behaved.  We present corresponding estimators for the cases of uncorrelated and correlated samples.
Harmonic mean estimators provide an estimation of the reciprocal of the marginal likelihood.  We therefore also consider estimation of the marginal likelihood itself and its variance from the reciprocal estimators.  Moreover, we present expressions to also estimate the Bayes factor, and its variance, to compare two models.
Finally, we present models to learn the normalised target distribution $\varphi(\theta)$ by approximating the posterior distribution, with the constraint that the target has narrower tails than the posterior, and discuss how to train such models.  Training involves constructing objective functions that penalise models that would result in estimators with a large variance, with appropriate regularisation.

%-------------------------------------------------------------------------------
\subsection{Uncorrelated samples}
%-------------------------------------------------------------------------------

MCMC algorithms that are typically used to sample the posterior distribution result in correlated samples.  By suitably thinning the MCMC chain (discarding all but every $t$th sample), however, samples that are uncorrelated can be obtained.  In this subsection we present estimators for the reciprocal marginal likelihood and its variance under the assumption of uncorrelated samples from the posterior.

Consider the harmonic moments
\begin{equation}
  \mu_n = \mathbb{E}_{\text{P}(\theta \given \data)} \Biggl[
    \biggl(\frac{\varphi(\theta)}{\mathcal{L}(\theta) \pi(\theta)}\biggr)^n
    \Biggr]
  \spcend ,
\end{equation}
and corresponding central moments
\begin{equation}
  \mu_n^\prime = \mathbb{E}_{\text{P}(\theta \given \data)} \Biggl[
    \biggl(
    \frac{\varphi(\theta)}{\mathcal{L}(\theta) \pi(\theta)}
    - \mathbb{E}_{\text{P}(\theta \given \data)}
    \biggl(\frac{\varphi(\theta)}{\mathcal{L}(\theta)
      \pi(\theta)}
    \biggr)
    \biggr)^n
    \Biggr]
  \spcend .
\end{equation}
We make use of the following harmonic moment estimators computed from samples of the posterior:
\begin{equation}
  \hat{\mu}_n =
  \frac{1}{N} \sum_{i=1}^N
  \biggl(\frac{\varphi(\theta_i)}{\mathcal{L}(\theta_i) \pi(\theta_i)}\biggr)^n \spcend ,
  \quad
  \theta_i \sim \text{P}(\theta \given \data)
  \spcend ,
\end{equation}
which are unbiased estimators of $\mu_n$, i.e. $\mathbb{E}(\hat{\mu}_n) = \mu_n$.
The reciprocal marginal likelihood can then be estimated from samples of the posterior by
\begin{equation}
  \hat{\rho} =
  \hat{\mu}_1 =
  \frac{1}{N} \sum_{i=1}^N
  \frac{\varphi(\theta_i)}{\mathcal{L}(\theta_i) \pi(\theta_i)} \spcend ,
  \quad
  \theta_i \sim \text{P}(\theta \given \data)
  \spcend .
\end{equation}
The mean and variance of the estimator read, respectively,
\begin{equation}
  \mathbb{E}(\hat{\rho})
  = \mathbb{E}
  \Biggl [
    \frac{1}{N} \sum_{i=1}^N
    \frac{\varphi(\theta_i)}{\mathcal{L}(\theta_i) \pi(\theta_i)}
    \Biggr ]
  = \mu_1
  = \rho
\end{equation}
and
\begin{equation}
  \label{eqn:var_rho_hat}
  \text{var}(\hat{\rho})
  = \text{var}
  \Biggl [
    \frac{1}{N} \sum_{i=1}^N
    \frac{\varphi(\theta_i)}{\mathcal{L}(\theta_i) \pi(\theta_i)}
    \Biggr ]
  = \frac{1}{N} (\mu_2 - \mu_1 ^2 )
  \spcend .
\end{equation}
Note that the estimator is unbiased.

Recall from \sectn{\ref{sec:review}} that the optimal target is given by the normalised posterior, i.e.\ $\varphi^\text{optimal}(\theta) = \mathcal{L}(\theta) \pi(\theta)/z$.  It is straightforward to see that in this case
\begin{equation}
  \mu_n
  = \hat{\mu}_n
  = \frac{1}{z^n}
  \spcend ,
\end{equation}
and thus the target distribution is optimal since
\begin{equation}
  \text{var}(\hat{\rho})
  = \frac{1}{N} (\mu_2 - \mu_1 ^2 )
  = \frac{1}{N} (1/z^2 - (1/z) ^2 )
  = 0
  \spcend .
\end{equation}

We are interested in not only an estimate of the reciprocal marginal likelihood but also its variance $\text{var}(\hat{\rho})$.  It is clear from \eqn{\ref{eqn:var_rho_hat}} that a suitable estimator of the variance is given by
\begin{equation}
  \hat{\sigma}^2 = \frac{1}{N-1} (\hat{\mu}_2 - \hat{\mu}_1 ^2 )
  =
  \frac{1}{N(N-1)} \sum_{i=1}^N
  \biggl(\frac{\varphi(\theta_i)}{\mathcal{L}(\theta_i) \pi(\theta_i)}\biggr)^2
  - \frac{\hat{\rho}^2}{N-1}
  \spcend .
\end{equation}
It follows that this estimator of the variance is unbiased since
\begin{equation}
  \mathbb{E}(\hat{\sigma}^2)
  = \frac{1}{N} (\mu_2 - \mu_1^2)
  = \text{var}(\hat{\rho})
  \spcend .
\end{equation}
The variance of the estimator $\hat{\sigma}^2$ reads
\begin{equation}
  \text{var}(\hat{\sigma}^2)
  = \frac{1}{(N-1)^2}
  \biggl[
    \frac{(N - 1)^2}{N^3} \mu_4^\prime - \frac{(N-1)(N-3)}{N^3} \mu_2^\prime{}^2
    \biggr]
  \spcend ,
\end{equation}
where $\mu_n^\prime$ are central moments, which follows by a well-known result for the variance of a sample variance \citep[e.g.][p.~264]{rose:2002}. An unbiased estimator of $\text{var}(\hat{\sigma}^2)$ can be constructed from h-statistics \citep[e.g.][]{rose:2002}, which provide unbiased estimators of central moments.

While we have presented general estimators for uncorrelated samples here, generating uncorrelated samples requires thinning the MCMC chain, which is highly inefficient.  It is generally recognised that thinning should be avoided when possible since it reduces the precision with which summaries of the MCMC chain can be computed \citep{link:2012}. Subsequently, we consider estimators that do not require uncorrelated samples and so can make use of considerably more MCMC samples of the posterior.

%-------------------------------------------------------------------------------
\subsection{Correlated samples}
\label{sec:learnt_harmonic_mean:correlated_samples}
%-------------------------------------------------------------------------------

We present an estimator of the reciprocal marginal likelihood, an estimate of the variance of this estimator, and its variance.  These estimators make use of correlated samples in order to avoid the loss of efficiency that results from thinning an MCMC chain.

We propose running a number of independent MCMC chains and using all of the correlated samples within a given chain.  A number of modern MCMC sampling techniques, such as affine invariance ensemble samplers \citep{goodman:2010}, naturally provide samples from multiple chains by their ensemble nature.  Moreover, excellent software implementations are readily available, such as the \texttt{emcee} code\footnote{\url{https://emcee.readthedocs.io/en/stable/}} \citep{foreman-mackey:2013}, which provides an implementation of the affine invariance ensemble samplers proposed by \citet{goodman:2010}.  Alternatively, if only a single large chain is available then this can be broken into separate blocks, which are (approximately) independent for a suitably long block length.  Subsequently, we use the terminology chains throughout to refer to both scenarios of running multiple MCMC chains or separating a single chain in blocks.

Consider $C$ chains of samples, indexed by $j = 1, 2, \ldots, C$, with chain $j$ containing $N_j$ samples.  The $i$th sample of chain $j$ is denoted $\theta_{ij}$.  Since the chain of interest is typically clear from the context, for notational brevity we drop the chain index from the samples, i.e.\ we denote samples by $\theta_i$ where the chain of interest is inferred from the context.

An estimator of the reciprocal marginal likelihood can be computed from each independent chain by
\begin{equation}
  \hat{\rho}_j =
  \frac{1}{N_j} \sum_{i=1}^{N_j}
  \frac{\varphi(\theta_i)}{\mathcal{L}(\theta_i) \pi(\theta_i)} \spcend ,
  \quad
  \theta_i \sim \text{P}(\theta \given \data)
  \spcend .
\end{equation}
A single estimator of the reciprocal marginal likelihood can then be constructed from the estimator for each chain by
\begin{equation}
  \hat{\rho}
  = \frac{\sum_{j=1}^{C} w_j \hat{\rho}_j}
  {\sum_{j=1}^{C} w_j }
  \spcend ,
\end{equation}
where the estimator $\hat{\rho}_j$ of chain $j$ is weighted by the number of samples in the chain, i.e.\ $w_j = N_j$.  It is straightforward to see that the estimator of the reciprocal marginal likelihood is unbiased, i.e.\ $\mathbb{E}(\hat{\rho})= \rho$, since $\mathbb{E}(\hat{\rho}_j) = \rho$.

The variance of the estimator $\hat{\rho}$ is related to the population variance $\sigma^2 = \mathbb{E}\bigl[ (\hat{\rho}_i - \mathbb{E}(\hat{\rho}_i))^2 \bigr]$ by
\begin{equation}
  \text{var}(\hat{\rho}) = \frac{\sigma^2}{N_\text{eff}}
  \spcend ,
\end{equation}
where the effective sample size is given by
\begin{equation}
  N_\text{eff} = \frac{\bigl(\sum_j^{C} w_j \bigr)^2}{\sum_j^{C} w_j^2}
  \spcend .
\end{equation}
The estimator of the population variance, given by
\begin{equation}
  \hat{s}^2
  = \frac{N_\text{eff}}
  {N_\text{eff}-1}
  \frac{\sum_{j=1}^{C} w_j (\hat{\rho}_j-\hat{\rho})^2}{\sum_j^{C} w_j}
  \spcend ,
\end{equation}
is unbiased, i.e.\ $\mathbb{E}(\hat{s}^2) = \sigma^2$.
A suitable estimator for $\text{var}(\hat{\rho})$ is thus
\begin{equation}
  \hat{\sigma}^2
  = \frac{\hat{s}^2}{N_\text{eff}}
  = \frac{1}
  {N_\text{eff}-1}
  \frac{\sum_{j=1}^{C} w_j (\hat{\rho}_j-\hat{\rho})^2}{\sum_j^{C} w_j}
  \spcend ,
\end{equation}
which is unbiased, i.e.\ $\mathbb{E}(\hat{\sigma}^2) = \text{var}(\hat{\rho})$, since $\hat{s}^2$ is unbiased.

The variance of the estimator $\hat{\sigma}^2$ reads
\begin{equation}
  \text{var}(\hat{\sigma}^2) = \frac{1}{N_\text{eff}{}^2} \text{var}(\hat{s}^2)
  = \frac{\sigma^4}{N_\text{eff}{}^3}
  \biggl(\kappa - 1 + \frac{2}{N_\text{eff}-1}\biggr)
  \spcend ,
\end{equation}
where in the second equality we have used a well-known result for the variance of the sample variance of independent and identically distributed (i.i.d.) random variables \citep[e.g.][]{cho:2005}.
The kurtosis $\kappa$ is defined by
\begin{equation}
  \kappa
  = \text{kur}(\hat{\rho}_i)
  = \mathbb{E} \Biggl[ \biggl(\frac{\hat{\rho}_i - \rho}{\sigma}\biggr)^4 \Biggr]
  \spcend .
\end{equation}
A suitable estimator for $\text{var}(\hat{\sigma}^2)$ is thus
\begin{align}
  \hat{\nu}^4 = \frac{\hat{s}^4}{N_\text{eff}{}^3}
  \biggl(\hat{\kappa} - 1 + \frac{2}{N_\text{eff}-1}\biggr)
  = \frac{\hat{\sigma}^4}{N_\text{eff}{}}
  \biggl(\hat{\kappa} - 1 + \frac{2}{N_\text{eff}-1}\biggr)
  \spcend ,
\end{align}
where for the kurtosis we adopt the estimator
\begin{align}
  \hat{\kappa}
  =
  \frac{\sum_{j=1}^{C} w_j (\hat{\rho}_j-\hat{\rho})^4}
  {\hat{s}^4 \sum_{j=1}^{C} w_j}
  =
  \frac{\sum_{j=1}^{C} w_j (\hat{\rho}_j-\hat{\rho})^4}
  {N_\text{eff}{}^2 \hat{\sigma}^4 \sum_{j=1}^{C} w_j}
\end{align}
(although alternative estimators of the kurtosis may by considered).

The estimators $\hat{\rho}$, $\hat{\sigma}^2$ and $\hat{\nu}^4$ provide a strategy to estimate the reciprocal marginal likelihood, its variance, and the variance of the variance, respectively.  The variance estimators provide valuable measures of the accuracy of the estimated reciprocal marginal likelihood and provide useful sanity checks.

Additional sanity checks can also be considered.
By the central limit theorem, for a large number of samples the distribution of $\hat{\rho}_j$ approaches a Gaussian, with kurtosis $\kappa=3$.  If the estimated kurtosis $\hat{\kappa} \gg 3$ it would indicate that the sampled distribution of $\hat{\rho}_j$ has long tails, suggesting further samples need to be drawn.
Similarly, the ratio of $\hat{\nu}^2 / \hat{\sigma}^2$ can be inspected to see if it is close to that expected for a Gaussian distribution with $\kappa=3$ of
\begin{equation}
  \frac{\hat{\nu}^4}{\hat{\sigma}^4} = \frac{1}{N_\text{eff}{}}
  \biggl(2 + \frac{2}{N_\text{eff}-1}\biggr)
  = \frac{2}{N_\text{eff}-1}
  \spcend ,
\end{equation}
or equivalently
\begin{equation}
  \frac{\hat{\nu}^2}{\hat{\sigma}^2}
  = \sqrt{\frac{2}{N_\text{eff}-1}}
  \spcend .
\end{equation}
For the common setting where the number of samples per chain is constant, i.e.\ $N_j = N$ for all $j$,
\begin{equation}
  N_\text{eff} = \frac{\bigl(\sum_j^{C} w_j \bigr)^2}{\sum_j^{C} w_j^2} = \frac{(N C)^2}{N^2 C} = C
\end{equation}
and, say $C=100$, we find
\begin{equation}
  \frac{\hat{\nu}^2}{\hat{\sigma}^2} = 0.14
  \spcend.
\end{equation}
In this setting significantly larger values of this ratio would suggest that further samples need to be drawn.

%-------------------------------------------------------------------------------
\subsection{Bayes factors}
\label{sec:learnt_harmonic_mean:bayes_factors}
%-------------------------------------------------------------------------------

We have so far considered the estimation of the reciprocal marginal likelihood and related variances only.  However it is the marginal likelihood itself (not its reciprocal), or the Bayes factors computed to compare two models, that is typically of direct interest.  We therefore consider how to compute these quantities of interest and a measure of their variance.

First, consider the mean and variance of the function $f(X,Y) = X/Y$ of two uncorrelated random variables $X$ and $Y$, which by Taylor expansion to second order are given by
\begin{equation}
  \mathbb{E}\biggl(\frac{X}{Y}\biggr) \simeq \frac{\mathbb{E}(X)}{\mathbb{E}(Y)} + \frac{\mathbb{E}(X)}{\mathbb{E}(Y)^3} \sigma_Y^2
\end{equation}
and
\begin{equation}
  \text{var}\biggl(\frac{X}{Y}\biggr) \simeq \frac{1}{\mathbb{E}(Y)^2} \sigma_X^2 + \frac{\mathbb{E}(X)^2}{\mathbb{E}(Y)^4} \sigma_Y^2
  \spcend ,
\end{equation}
respectively, where $\sigma_X = \mathbb{E}\bigl[ (X - \mathbb{E}(X))^2 \bigr]$ and $\sigma_Y = \mathbb{E}\bigl[ (Y - \mathbb{E}(Y))^2 \bigr]$.

Using this result the marginal likelihood and its variance can be estimated from the reciprocal estimators by making use of the relations
\begin{equation}
  \mathbb{E}( z )
  = \mathbb{E}\biggl(\frac{1}{{\rho}}\biggr)
  \simeq \frac{1}{\mathbb{E}({\rho})} \biggl( 1 + \frac{\sigma_{\rho}^2}{\mathbb{E}({\rho})^2} \biggr)
  %= \hat{z}
\end{equation}
and
\begin{equation}
  \text{var}\biggl(\frac{1}{{\rho}}\biggr)
  \simeq \frac{\sigma_{\rho}^2}{\mathbb{E}({\rho})^4}
  %= \hat{\epsilon}^2
  \spcend ,
\end{equation}
respectively, by considering the case $X=1$ and $Y = \rho$.

Typically it is the Bayes factor given by the ratio of marginal likelihoods that is of most interest in order to compare models.  Again using the expressions above for the mean and variance of the function $f(X,Y) = X/Y$, this time for the case $X = \rho_2$ and $Y = \rho_1$, the Bayes factor and its variance can be estimated directly from the reciprocal marginal likelihood estimates and variances by making use of the relations
\begin{equation}
  \mathbb{E}\biggl(\frac{z_1}{z_2}\biggr)
  =
  \mathbb{E}\biggr(\frac{\rho_2}{\rho_1}\biggr)
  \simeq
  \frac{\mathbb{E}({\rho_2})}{\mathbb{E}({\rho_1})}
  \biggl( 1 + \frac{\sigma_{\rho_1}^2}{\mathbb{E}({\rho_1})^2}  \biggr)
  %= \hat{z}_{12}
\end{equation}
and
\begin{equation}
  \text{var}\biggl(\frac{z_1}{z_2}\biggr)
  =
  \text{var}\biggl(\frac{{\rho_2}}{{\rho_1}}\biggr)
  \simeq
  % \frac{1}{\mathbb{E}({\rho_1})^2} \sigma_{\rho_2}^2 + \frac{\mathbb{E}({\rho_2})^2}{\mathbb{E}({\rho_1})^4} \sigma_{\rho_1}^2
  % =
  \frac{\mathbb{E}({\rho_1})^2 \sigma_{\rho_2}^2+ \mathbb{E}({\rho_2})^2 \sigma_{\rho_1}^2}{\mathbb{E}({\rho_1})^4}
  %= \hat{\epsilon}_{12}^2
  \spcend ,
\end{equation}
respectively.

%-------------------------------------------------------------------------------
\subsection{Learning the target density}
\label{sec:learnt_harmonic_mean:target}
%-------------------------------------------------------------------------------

While we have described estimators to compute the marginal likelihood and Bayes factors based on a learnt target distribution $\varphi(\theta)$, we have yet to consider the critical task of learning the target distribution.  As discussed, the ideal target distribution is the posterior itself.  However, since the target must be normalised, use of the posterior would require knowledge of the marginal likelihood -- precisely the quantity that we attempting to estimate.  Instead,
one can learn an approximation of the posterior that is normalised.  The approximation itself does not need to be highly accurate.  More critically, the learned target approximating the posterior must exhibit narrower tails than the posterior to avoid the problematic scenario of the original harmonic mean that can result in very large variance.

We present three examples of models that can be used to learn appropriate target distributions and discuss how to train them, although other models can of course be considered.  Samples of the posterior are split into training and evaluation (cf.\ test) sets.  The training set is used to learn the target distribution, after which the evaluation set, combined with the learnt target, is used to estimate the marginal likelihood.  To train the models we typically construct and solve an optimisation problem to minimise the variance of the estimator, while ensuring it is unbiased.  We typically solve the resulting optimisation problem by stochastic gradient descent.  To set hyperparameters, we advocate cross-validation.

%-------------------------------------------------------------------------------
\subsubsection{Hypersphere}

The simplest model one may wish to consider is a hypersphere, much like the truncated harmonic mean estimator.  However, here we learn the optimal radius of the hypersphere, rather than setting the radius based on arbitrary level-sets of the posterior as considered previously.

Consider the target distribution defined by the normalised hypersphere
\begin{equation}
  \varphi(\theta) = \frac{1}{V_\mathcal{S}}  I_\mathcal{S}(\theta)
  \spcend ,
\end{equation}
where the indicator function $I_\mathcal{S}(\theta)$ is unity if $\theta$ is within a hypersphere of radius $R$, centred on $\bar{\theta}$ with covariance $\Sigma$, i.e.
\begin{equation}
  I_\mathcal{S}(\theta) =
  \begin{cases}
    1, & \bigl(\theta - \bar{\theta}\bigr)^\text{T} \Sigma^{-1} \bigl(\theta - \bar{\theta} \bigr) < R^2 \\
    0, & \text{otherwise}
  \end{cases}
  \spcend .
\end{equation}
The values of $\bar{\theta}$ and $\Sigma$ can be computed directly from the training samples.  Often, although not always, a diagonal approximation of $\Sigma$ is considered for computational efficiency.
The volume of the hypersphere required to normalise the distribution is given by
\begin{equation}
  V_\mathcal{S} = \frac{\pi^{d/2}}{\Gamma(d/2 + 1)} R^d \, \vert \Sigma \vert^{1/2}
  \spcend .
\end{equation}
Recall that $d$ is the dimension of the parameter space, i.e.\ $\theta \in \mathbb{R}^d$, and note that $\Gamma(\cdot)$ is the Gamma function.

To estimate the radius of the hypersphere we pose the following optimisation problem to minimise the variance of the learnt harmonic mean estimator, while also constraining it be be unbiased:
\begin{equation}
  \min_R \: \hat{\sigma}^2
  \quad \text{s.t.} \quad \hat{\rho} = \hat{\mu}_1
  \spcend .
\end{equation}
By minimising the variance of the estimator we ensure, on one hand, that the tails of the learnt target are not so wide that they are broader than the posterior, and, on the other hand, that they are not so narrow that very few samples are effectively retained in the estimator.
This optimisation problem is equivalent to minimising the estimator of the second harmonic moment:
\begin{equation}
  \min_R \: \hat{\mu}_2
  \spcend .
\end{equation}
Writing out the cost function explicitly in terms of the posterior samples, the optimisation problem reads
\begin{equation}
  \min_R \: \sum_i C_i^2
  \spcend ,
\end{equation}
with costs for each sample given by
\begin{equation}
  C_i
  = \frac{\varphi(\theta_i)}{\mathcal{L}(\theta_i) \pi(\theta_i)}
  \propto
  \begin{cases}
    \frac{1}{\mathcal{L}(\theta_i) \pi(\theta_i) R^d}, & \bigl(\theta - \bar{\theta}\bigr)^\text{T} \Sigma^{-1} \bigl(\theta - \bar{\theta} \bigr) < R^2 \\
    0,                                                 & \text{otherwise}
  \end{cases}
  \spcend .
\end{equation}
This one-dimensional optimisation problem can be solved by straightforward techniques, such as the Brent hybrid root-finding algorithm.

While the learnt hypersphere model is very simple, it is good pedagogical illustration of the general procedure for learning target distributions.  First, construct a normalised model.  Second, train the model to learn its parameters by solving an optimisation problem to minimise the variance of the estimator while ensuring it is unbiased.  If required, set hyperparameters or compare alternative models by cross-validation.
While the simple learnt hypersphere model may be sufficient in some settings, it is not effective for multimodal posterior distributions or for posteriors with narrow curving degeneracies.  For such scenarios we consider alternative learnt models.

%-------------------------------------------------------------------------------
\subsubsection{Modified Gaussian mixture model}

A modified Gaussian mixture model provides greater flexibility that the simple hypersphere model.  In particular, it is much more effective for multimodal posterior distributions.

Consider the target distribution defined by the modified Gaussian mixture model
\begin{equation}
  \varphi(\theta) = \sum_{k=1}^K \frac{w_k}{(2\pi)^{d/2} \vert \Sigma_k \vert^{1/2} s_k^d}
  \exp \biggl(
  \frac{- \bigl(\theta - \bar{\theta}_k\bigr)^\text{T} \Sigma_k^{-1} \bigl(\theta - \bar{\theta}_k\bigr)}
  {2 s_k^2}
  \biggr)
  \spcend ,
\end{equation}
for $K$ components, with centres $\bar{\theta}_k$ and covariances $\Sigma_k$, where the relative scale of each component is controlled by $s_k$ and the weights are specified by
\begin{equation}
  w_k = \frac{\exp(z_k)}{\sum_{k^\prime=1}^K \exp(z_{k^\prime})}
  \spcend ,
\end{equation}
which in turn depend on the weights $z_k$.  Given $K$, the posterior training samples can be clustered by $K$-means.  The values of $\bar{\theta}_k$ and $\Sigma_k$ can then be computed by the samples in cluster $k$.  The model is modified relative to the usual Gaussian mixture model in that the cluster mean and covariance are estimated from the samples of each cluster, while the relative cluster scale and weights are fitted.  Moreover, as before, a bespoke training approach is adopted tailored to the problem of learning an effective model for the learnt harmonic mean estimator.

To estimate the the weights $z_k$, which in turn define the weights $w_k$, and the relative scales $s_k$ we again construct an optimisation problem to minimise the variance of the learnt harmonic mean estimator, while also constraining it to be unbiased.  We also regularise the relative scale parameters, resulting in the following optimisation problem:
\begin{equation}
  \min_{\{z_k,s_k\}_{k=1}^K} \: \hat{\sigma}^2 + \frac{1}{2} \lambda \sum_{k=1}^K s_k^2
  \quad \text{s.t.} \quad \hat{\rho} = \hat{\mu}_1
  \spcend ,
\end{equation}
for regularisation parameter $\lambda$.
The problem may equivalently be written as
\begin{equation}
  \min_{\{z_k,s_k\}_{k=1}^K}
  \: \hat{\mu}_2 + \frac{1}{2} \lambda \sum_{k=1}^K s_k^2
  \spcend ,
\end{equation}
or explicitly in terms of the posterior samples by
\begin{equation}
  \min_{\{z_k,s_k\}_{k=1}^K}
  \: \sum_i C_i^2 + \frac{1}{2} \lambda \sum_{k=1}^K s_k^2
  \spcend.
\end{equation}
The individual cost terms for each sample $i$ are given by
\begin{equation}
  C_i
  = \frac{\varphi(\theta_i)}{\mathcal{L}(\theta_i) \pi(\theta_i)}
  = \sum_{k=1}^K C_{ik}
  \spcend ,
\end{equation}
which include the following component from cluster $k$:
\begin{equation}
  C_{ik} = \frac{w_k}{(2\pi)^{d/2} \vert \Sigma_k \vert^{1/2} s_k^d}
  \,
  \exp \biggl(
  \frac{- \bigl(\theta_i - \bar{\theta}_k\bigr)^\text{T} \Sigma_k^{-1} \bigl(\theta_i - \bar{\theta}_k\bigr)}
  {2 s_k^2}
  \biggr)
  \frac{1}{\mathcal{L}(\theta_i) \pi(\theta_i)}
  \spcend .
\end{equation}

We solve this optimisation problem by stochastic gradient decent, which requires the gradients of the objective function.  Denoting the total cost of the objective function by $C = \sum_i C_i^2 + \frac{1}{2} \lambda \sum_{k=1}^K s_k^2$, it is straightforward to show that the gradients of the cost function with respective to the weights $z_k$ and relative scales $s_k$ are given by
\begin{equation}
  \frac{\partial C}{\partial z_k} = 2 \sum_i C_i (C_{ik} - w_k C_i)
\end{equation}
and
\begin{equation}
  \frac{\partial C}{\partial s_k} = 2 \sum_i \frac{C_i C_{ik}}{s_k^3}
  \Bigl(
  \bigl(\theta_i - \bar{\theta}_k\bigr)^\text{T} \Sigma_k^{-1} \bigl(\theta_i - \bar{\theta}_k\bigr)
  - d s_k^2
  \Bigr)
  \spcend ,
\end{equation}
respectively.

The general procedure to learn the target distribution is the same as before: first, construct a normalised model; second, train the model by solving an optimisation problem to minimise the variance of the resulting learnt harmonic mean estimator.  In this case we regularise the relative scale parameters and then solve by stochastic gradient descent.  The number of clusters $K$ can be deteremined by cross-validation (or other methods).
While the modified Gaussian mixture model can effectively handle multimodal distributions, alternative models are better suited to narrow curving posterior degeneracies.

%-------------------------------------------------------------------------------
\subsubsection{Kernel density estimation}

Kernel density estimation (KDE) provides another alternative model to learn an effective target distribution. In particular, it can be used to effectively model narrow curving posterior degeneracies.

Consider the target distribution defined by the kernel density function
\begin{equation}
  \varphi(\theta) = \frac{1}{N} \sum_i \frac{1}{V_{K}} K(\theta - \theta_i)
  \spcend ,
\end{equation}
with kernel
\begin{equation}
  K(\theta) = k\biggl(\frac{\theta^\text{T} \Sigma_K^{-1} \theta}{R^2} \biggr)
  \spcend ,
\end{equation}
where $k(\theta) = 1$ if $\vert \theta \vert < 1/2$ and 0 otherwise.  The volume of the kernel is given by
\begin{equation}
  V_{K} = \frac{\pi^{d/2}}{\Gamma(d/2+1)} R^d \vert \Sigma_K \vert^{1/2}
  \spcend .
\end{equation}
The kernel covariance $\Sigma_K$ can be computed directly from the training samples, for example by estimating the covariance or even simply by the separation between the lowest and highest samples in each dimension.  A diagonal representation is often, although not always, considered for computational efficiency.

The kernel radius $R$ can be estimating by following a similar procedure to those outlined above for the hypersphere and modified Gaussian mixture model to minimise the variance of the resulting estimator.  Alternatively, since there is only a single parameter cross-validation is also effective.

%===============================================================================
\section{Numerical experiments}
\label{sec:experiments}
%===============================================================================

We perform numerous numerical experiments to validate the learnt harmonic mean estimator by comparing to ground truth marginal likelihood values for a variety of example problems.  The techniques presented in \sectn{\ref{sec:learnt_harmonic_mean}} are implemented in the \texttt{harmonic} software package\footnote{\url{https://github.com/astro-informatics/harmonic}}, which is discussed further in \sectn{\ref{sec:code}}.  Throughout we use \texttt{harmonic} with the \texttt{emcee} code\footnote{\url{https://emcee.readthedocs.io/en/stable/}} \citep{foreman-mackey:2013} to perform MCMC sampling.   We consider problems with narrow curving posterior degeneracies, multimodal distributions, and scenarios where the original harmonic mean estimator has been shown to fail catastrophically, while applying all three of the strategies to learn the target density $\varphi(\theta)$ that are discussed in \sectn{\ref{sec:learnt_harmonic_mean:target}}.  In all cases the learnt harmonic mean estimator is shown to be robust and highly accurate.

%-------------------------------------------------------------------------------
\subsection{Rosenbrock}
%-------------------------------------------------------------------------------

A common benchmark problem to test methods to compute the marginal likelihood is a likelihood specified by the Rosenbrock function.  The Rosenbrock function exhibits a narrow curving degeneracy, which makes it challenging to explore the resulting posterior sufficiently to evaluate the marginal likelihood accurately.

The Rosenbrock function is given by
\begin{equation}
  f({x}) = \sum_{i=1}^{d-1} \bigg [ 100(x_{i+1} - x_{i}^2)^2 + (x_i - 1)^2 \bigg ]
  \spcend ,
\end{equation}
where $d$ denotes dimension.
Due to its very narrow curving degeneracy, it can be difficult to numerically estimate the minimum of the Rosenbrock function, which can be seen analytically is given by $f({x}_{\text{min}}) = 0$ at ${x}_{\text{min}} = (1,\dots,1)$.
We consider a log-likelihood given by $\log \mathcal{L}(x) = -f({x})$ and consider a simple uniform prior with $x_0 \in [-10, 10]$ and $x_1 \in [-5, 15]$.

\begin{figure}
  \centering
  \includegraphics[width=0.45\textwidth]{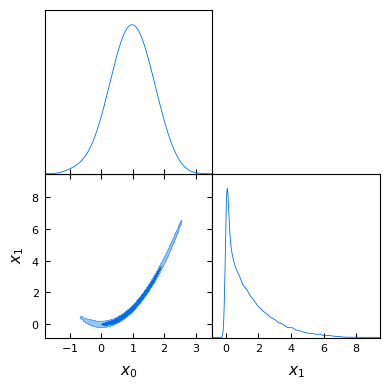}
  \caption{Rosenbrock posterior recovered by MCMC sampling using \texttt{emcee}. The Rosenbrock function exhibits a narrow curving degeneracy, which makes it challenging to explore the resulting posterior sufficiently to evaluate the marginal likelihood accurately.}
  \label{fig:rosenbrock_posterior}
\end{figure}

We compute the marginal likelihood for dimension $d=2$ using the learnt harmonic mean estimator and by brute force to provide a ground truth for comparison, evaluating the marginal likelihood by numerical integration (which is possible in this low-dimensional setting).
For our learnt harmonic mean estimator, computed using \texttt{harmonic}, we sample the resulting posterior distribution using \texttt{emcee}, drawing 5,000 samples for 200 chains, with burn in of 2,000 samples, yielding 3,000 posterior samples per chain.
The recovered posterior distribution is illustrated in \fig{\ref{fig:rosenbrock_posterior}}.
We use 50\% of the samples to fit a KDE model for the target distribution (recall that the KDE model is well-suited to problems with narrow curving degeneracies), using cross-validation to estimate the model hyperparameters.  The remaining 50\% of posterior samples are used to infer the marginal likelihood.  Computation time is about one minute to compute on a standard laptop, including drawing all samples and performing cross-validation.  We repeat this experiment 100 times in order to estimate the variance of the estimator and its variance, in order to compare to the variance and variance-of-variance estimators described in \sectn{\ref{sec:learnt_harmonic_mean:correlated_samples}}.

\begin{figure}
  \centering
  \subfigure[Inverse evidence]{\includegraphics[width=0.45\textwidth]{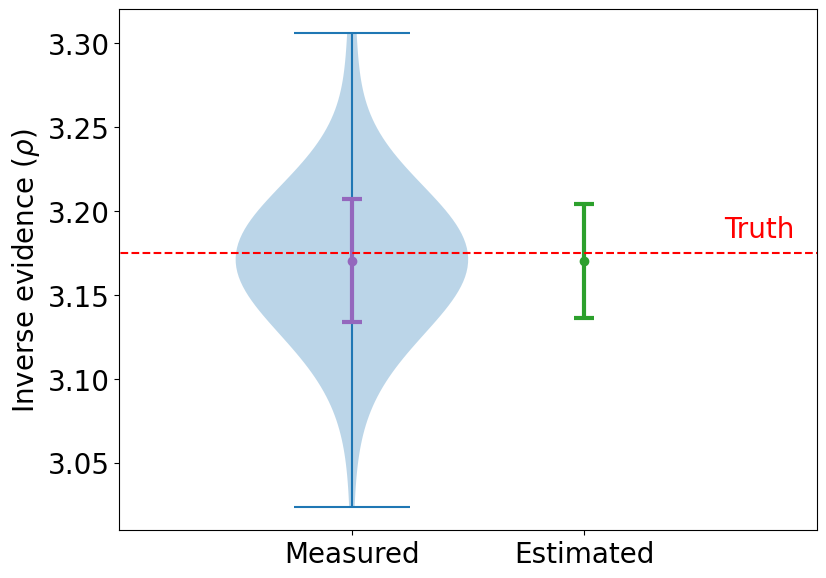}} \quad
  \subfigure[Variance of inverse evidence]{\includegraphics[width=0.45\textwidth]{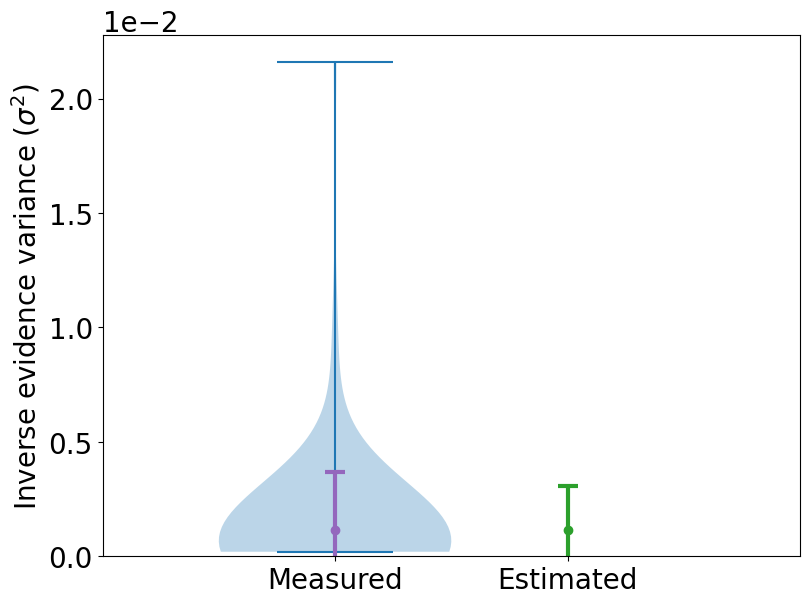}}
  \caption{Marginal likelihood (evidence) computed by the learnt harmonic mean estimator (using \texttt{harmonic}) for the Rosenbrock benchmark problem.  100 experiments are repeated to recover empirical estimates of the statistics of the estimator.  In panel~(a) the distribution of marginal likelihood values computed by the learnt harmonic mean estimator are shown, with the mean and standard deviation of the distribution also shown (measured).  For comparison the estimate of the standard deviation computed by the error estimator is also shown (estimated).  The ground truth estimated by numerical integration is indicated by the red dashed line.  In panel~(b) the distribution of the variance estimator is shown, with its mean and standard deviation (estimated).  For comparison the standard deviation computed by the variance-of-variance estimator is also shown (estimated).  The learnt harmonic mean estimator and its error estimators are highly accurate.}
  \label{fig:rosenbrock_evidence}
\end{figure}

The distribution of marginal likelihood values compute by our learnt harmonic mean estimator for all 100 experiments are shown in \fig{\ref{fig:rosenbrock_evidence}}.  In addition, we show the values computed by the variance and variance-of-variance estimators (estimated) and compare them to the corresponding statistics measured from the 100 experiments (measured).  Moreover, we plot the ground truth value computed by numerical integration.  The marginal likelihood value computed is in close agreement with the ground truth and the variance and variance-of-variance estimators are in close agreement with the values computed from the experiments.  It is clear that the learnt harmonic mean estimator is highly accurate, its variance is well-behaved and its error estimators are also highly accurate.

%-------------------------------------------------------------------------------
\subsection{Rastrigin}
%-------------------------------------------------------------------------------

Another common benchmark problem to test marginal likelihood estimators is a likelihood specified by the Rastrigin function.  The Rastrigin function exhibits multiple local peaks, which makes it challenging to explore the resulting posterior sufficiently to evaluate the marginal likelihood accurately.

The Rastrigin function is given by
\begin{equation}
  f({x}) = 10 d + \sum_{i=1}^{d} \bigl [ x_i^2 - 10 \cos ( 2 \pi x_i ) \bigr ]
  \spcend ,
\end{equation}
where $d$ denotes dimension.
Due to its highly multimodal behaviour, it can be difficult to numerically estimate the minimum of the Rastrigin function.  Its local minima are given by integer coordinate values, with the global minimum at \mbox{${x}_\text{min} = 0$}.
We consider a log-likelihood given by $\log \mathcal{L}(x) = -f({x})$ and consider a simple uniform prior with $x_i \in [-6, 6]$ for $i = 1, \dots, d$.

\begin{figure}
  \centering
  \includegraphics[width=0.45\textwidth]{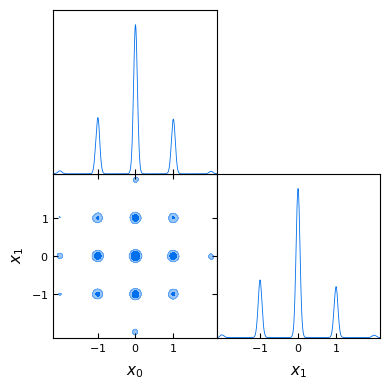}
  \caption{Rastrigin posterior recovered by MCMC sampling using \texttt{emcee}.  The Rastrigin function exhibits multiple local peaks, which makes it challenging to explore the resulting posterior sufficiently to evaluate the marginal likelihood accurately.}
  \label{fig:rastrigin_posterior}
\end{figure}

We compute the marginal likelihood for dimension $d=2$ in an identical manner as for the Rosenbrock example, that is, using the learnt harmonic mean estimator and by brute force to provide a ground truth for comparison, evaluating the marginal likelihood by numerical integration (which, again, is possible in this low-dimensional setting).
For our learnt harmonic mean estimator, computed using \texttt{harmonic}, we sample the resulting posterior distribution using \texttt{emcee}, drawing 5,000 samples for 200 chains, with burn in of 2,000 samples, yielding 3,000 posterior samples per chain.
The recovered posterior distribution is illustrated in \fig{\ref{fig:rastrigin_posterior}}.
We again adopt a KDE model for the target distribution, avoiding the need to estimate the number of modes in the distribution, and use 50\% of the samples to fit the model, using cross-validation to estimate the model hyperparameters.  The remaining 50\% of posterior samples are used to infer the marginal likelihood.  Computation time is about one minute to compute on a standard laptop, including drawing all samples and performing cross-validation.  We again repeat this experiment 100 times in order to estimate the variance of the estimator and its variance, in order to compare to the variance and variance-of-variance estimators.

\begin{figure}
  \begin{center}
    \subfigure[Inverse evidence $1/z$]{\includegraphics[width=.45\textwidth, trim=0cm 0cm 0cm 0cm, clip=true]{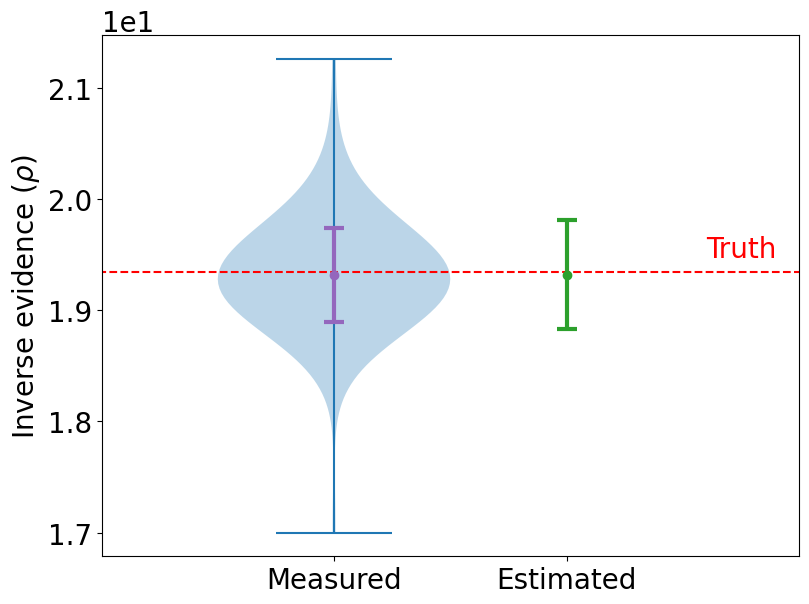}} \quad
    \subfigure[Variance on inverse evidence var$(1/z)$]{\includegraphics[width=.45\textwidth, trim=0cm 0cm 0cm 0cm, clip=true]{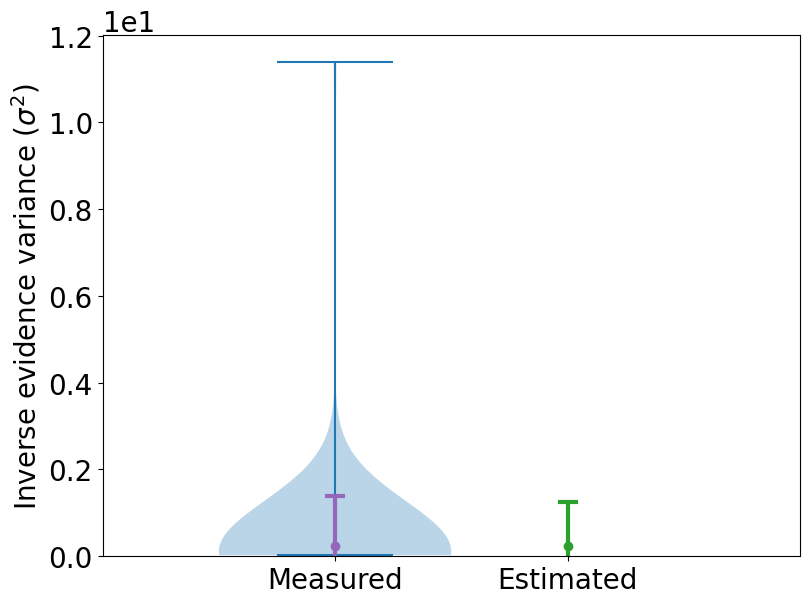}}
    \caption{Marginal likelihood (evidence) computed by the learnt harmonic mean estimator (using \texttt{harmonic}) for the Rastrigin benchmark problem (as in \fig{\ref{fig:rosenbrock_evidence}}).   100 experiments are repeated to recover empirical estimates of the statistics of the estimator.  In panel~(a) the distribution of marginal likelihood values computed by the learnt harmonic mean estimator are shown, with the mean and standard deviation of the distribution also shown (measured).  For comparison the estimate of the standard deviation computed by the error estimator is also shown (estimated).  The ground truth estimated by numerical integration is indicated by the red dashed line.  In panel~(b) the distribution of the variance estimator is shown, with its mean and standard deviation (estimated).  For comparison the standard deviation computed by the variance-of-variance estimator is also shown (estimated).  The learnt harmonic mean estimator and its error estimators are highly accurate.}
    \label{fig:rastrigin_evidence}
  \end{center}
\end{figure}

The distribution of marginal likelihood values computed by our learnt harmonic mean estimator for all 100 experiments are shown in \fig{\ref{fig:rastrigin_evidence}}.  As before, we also show the values computed by the variance and variance-of-variance estimators (estimated) and compare them to the corresponding statistics measured from the 100 experiments (measured).  Moreover, we plot the ground truth value computed by numerical integration.
It is again clear that the learnt harmonic mean estimator is highly accurate, its variance is well-behaved and its error estimators are also highly accurate.

%-------------------------------------------------------------------------------
\subsection{Normal-Gamma}
%-------------------------------------------------------------------------------

An analytically tractable numerical example is considered in \cite{friel:2012} to assess the sensitivity of marginal likelihood estimators to changes in the prior.  In this study \cite{friel:2012} found that the marginal likelihood values computed by the original harmonic mean estimator do not vary with the prior as the values computed analytically do, highlighting this example as a pathological failure of the original harmonic mean estimator.  We consider the same pathological example here and demonstrate that our learnt harmonic mean estimator is highly accurate.

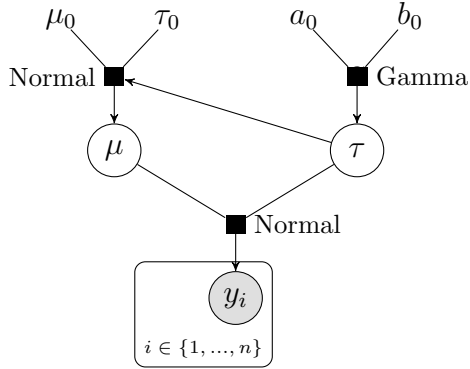
\begin{figure}
  \begin{center}
    % !TeX root = harmonic_cosmo21_2018.tex

\begin{tikzpicture}[scale=1.0, transform shape]
  [squarednode/.style={rectangle, draw=black, minimum size=8mm},
    latent/.style={circle, draw=black, minimum size=10mm}]

  % \footnotesize

  % Define nodes
  \node[obs]                               (y) {\large $y_i$};

  \node[factor, above=0.5cm of y, xshift=0cm]  (N2) {N};
  % \node[const, right=0.1cm of G] () {\normalsize $\mathcal{G}$};
  \node[const, right=0.1cm of N2] () {\normalsize Normal};

  \node[latent, above=1.25cm of y, xshift=1.6cm]  (tau) {\large $\tau$};
  \node[const, above=1.25cm of tau, xshift=-0.7cm]  (a0) {\large $a_0$};
  \node[const, above=1.25cm of tau, xshift=0.7cm]  (b0) {\large $b_0$};

  \node[factor, above=0.5cm of tau, xshift=0cm]  (G) {G};
  % \node[const, right=0.1cm of G] () {\normalsize $\mathcal{G}$};
  \node[const, right=0.1cm of G] () {\normalsize Gamma};

  \node[latent, above=1.25cm of y, xshift=-1.6cm]  (mu) {\large $\mu$};
  \node[const, above=1.25cm of mu, xshift=-0.7cm]  (mu0) {\large $\mu_0$};
  \node[const, above=1.25cm of mu, xshift=0.7cm]  (tau0) {\large $\tau_0$};

  \node[factor, above=0.5cm of mu, xshift=0cm]  (N) {N};
  % \node[const, right=0.1cm of G] () {\normalsize $\mathcal{G}$};
  \node[const, left=0.1cm of N] () {\normalsize Normal};

  % \draw [->] (tau) -- (y);
  \draw [->] (tau) -- (N);
  % \draw [->] (mu) -- (y);

  \factoredge{a0,b0}{G}{tau}
  \factoredge{mu0,tau0}{N}{mu}
  \factoredge{mu,tau}{N2}{y}

  \platenode {} {(y)} {$i \in \{1, ..., n \}$}; %

\end{tikzpicture}
    \caption{Graphical representation of the Normal-Gamma model.}
    \label{fig:hbm_normal_gamma}
  \end{center}
\end{figure}

We consider the \textit{Normal-Gamma model} \citep{bernardo:1994} with data
\begin{equation}
  y_i \sim \text{N}(\mu, \tau^{-1})
  \spcend ,
\end{equation}
for $i \in \{1, \ldots, n\}$, with mean $\mu$ and precision (inverse variance) $\tau$.  A normal prior is assumed for $\mu$ and a Gamma prior for $\tau$:
\begin{align}
  \mu  & \sim \text{N}\bigl(\mu_0, (\tau_0 \tau)^{-1}\bigr) \spcend , \\
  \tau & \sim \text{Ga}(a_0, b_0) \spcend ,
\end{align}
with mean $\mu_0 = 0$, shape $a_0 = 10^{-3}$ and rate $b_0 = 10^{-3}$.  The precision scale factor $\tau_0$ is varied to observe the impact of changing prior on the computed marginal likelihood.
The joint prior for $(\mu, \tau)$ then reads:
\begin{align}
  \pi(\mu, \tau) & = \pi(\mu \given \tau) \pi(\tau) \\
                 & =
  \frac{b_0^{a_0} \sqrt{\tau_0}}{\Gamma(a_0) \sqrt{2 \pi}}
  \tau^{a_0-1/2}
  \exp(-b_0 \tau)
  \exp\bigl(-\tau_0 \tau(\mu-\mu_0)^2/2\bigr)
  \spcend .
\end{align}
The likelihood is given by
\begin{align}
  \mathcal{L}(y)
   & = \prod_{i=1}^n \text{P}(y_i \given \mu, \tau)             \\*
   & = \prod_{i=1}^n \sqrt{\frac{\tau}{2\pi}}
  \exp\Bigl(- \frac{\tau}{2} (y_i - \mu)^2\Bigr)                \\
   & = \Bigl(\frac{\tau}{2\pi}\Bigr)^{n/2}
  \exp\biggl(- \frac{\tau}{2} \sum_{i=1}^n (y_i - \mu)^2\biggr) \\
   & = \Bigl(\frac{\tau}{2\pi}\Bigr)^{n/2}
  \exp\Bigl(- \frac{\tau n}{2} \bigl(s^2 + (\bar{y} - \mu)^2\bigr)\Bigr)
  \spcend ,
\end{align}
where  $y = (y_1, \dots, y_n)^\text{T}$,
\begin{equation}
  \bar{y} =
  \frac{1}{n} \sum_{i=1}^n y_i
\end{equation}
and
\begin{equation}
  s^2 =
  \frac{1}{n} \sum_{i=1}^n (y_i - \bar{y})^2
  \spcend .
\end{equation}
A graphical representation of the Normal-Gamma model is illustrated in \fig{\ref{fig:hbm_normal_gamma}}.

For the Normal-Gamma model the marginal likelihood may be computed analytically by
\begin{equation}
  z =
  (2\pi)^{-n/2}
  \frac{\Gamma(a_n)}{\Gamma(a_0)}
  \frac{b_0^{a_0}}{b_n^{a_n}}
  \biggl(\frac{\tau_0}{\tau_n}\biggr)^{1/2}
  \spcend ,
\end{equation}
where
\begin{equation}
  \tau_n = \tau_0 + n
  \spcend ,
\end{equation}
\begin{equation}
  a_n = a_0 + n/2
\end{equation}
and
\begin{align}
  b_n
   & = b_0
  + \frac{1}{2} \sum_{i=1}^n (y_i - \bar{y})^2
  + \frac{\tau_0 n (\bar{y} - \mu_0)^2}{2(\tau_0+n)} \\*
   & = b_0
  + \frac{1}{2} n s^2
  + \frac{\tau_0 n (\bar{y} - \mu_0)^2}{2(\tau_0+n)}
  \spcend .
\end{align}

To assess the impact of altering the prior, we compute the marginal likelihood both analytically and using our learnt harmonic mean estimator for priors corresponding to $\tau_0 \in \{10^{-4}, 10^{-3}, 10^{-2}, 10^{-1}, 10^{0} \}$.  Data are simulated with underlying parameters $(\mu,\tau) = (0,1)$ to generate $n=100$ synthetic observations (the same experimental configuration considered by \citealt{friel:2012}).

\begin{table}
  \caption{Marginal likelihood values computed analytically and by the learnt harmonic mean estimator for the Normal-Gamma example.  While the original harmonic mean estimator fails catastrophically, our learnt harmonic mean estimator is highly accurate.}% ${}^*$Denotes computed by \cite{friel:2012}.}
  \label{tbl:normal_gamma_comparison}
  \centering
  \begin{tabular}{lrrrrr}\toprule
    $\tau_0$                  & \multicolumn{1}{c}{$10^{-4}$} & \multicolumn{1}{c}{$10^{-3}$} & \multicolumn{1}{c}{$10^{-2}$} & \multicolumn{1}{c}{$10^{-1}$} & \multicolumn{1}{c}{$10^{0}$} \\ \midrule
    Analytic $\log(z)$        & -144.5530                     & -143.4017                     & -142.2505                     & -141.0999                     & -139.9552                    \\
    Estimated $\log(\hat{z})$ & -144.5545                     & -143.3990                     & -142.2490                     & -141.1001                     & -139.9558                    \\
    Error                     & -0.0015                       & 0.0027                        & 0.0015                        & -0.0011                       & -0.0006                      \\
    {\scriptsize (learnt harmonic mean)}\hspace*{-10mm}                                                                                                                                      \\ \midrule
    Error                     & 12.2100                       & ---                           & 9.7900                        & 8.5000                        & 7.1000                       \\
    {\scriptsize (original harmonic mean)}\hspace*{-10mm}                                                                                                                                    \\ \bottomrule
  \end{tabular}
\end{table}

For our learnt harmonic mean estimator we use \texttt{emcee} to draw 1,500 samples for 200 chains, with burn in of 500 samples, yielding 1,000 posterior samples per chain.  We use 25\% of the samples to learn the target model, using cross-validation to select between the hypersphere and modified Gaussian mixture model.  In all cases the modified Gaussian mixture model is selected.  The remaining 75\% of posterior samples are used for inferring the marginal likelihood.  Computation time is about one minute on a standard laptop for each experiment (i.e.\ each $\tau_0$) considered, including drawing all samples.

\begin{figure}
  \centering \includegraphics[width=0.65\textwidth]{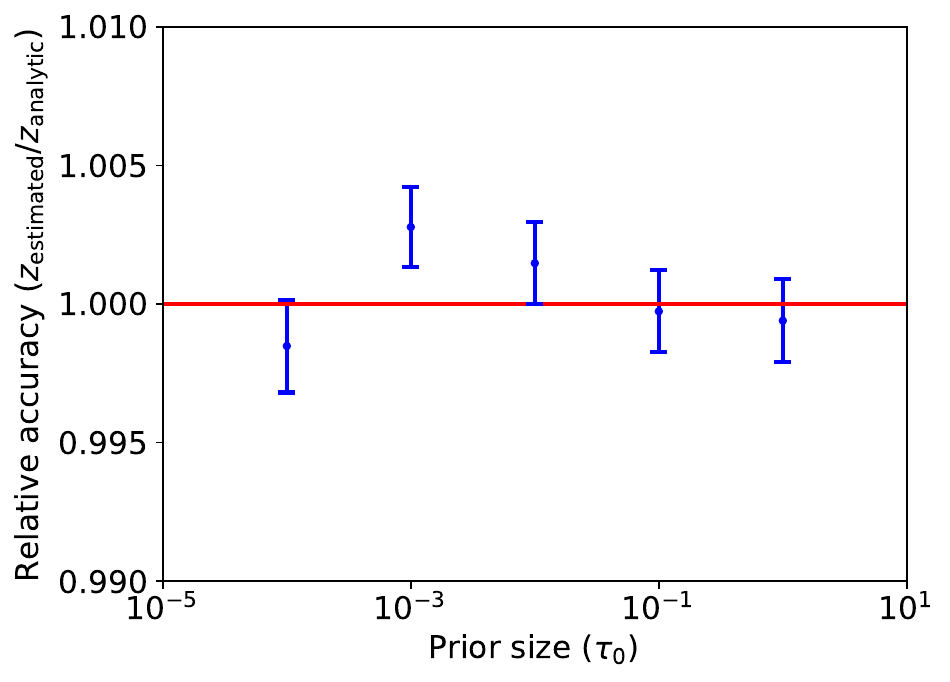}
  \caption{Ratio of marginal likelihood values computed by the learnt harmonic mean estimator to those computed analytically.  Errors bars corresponding to the estimated standard deviation of the learnt harmonic estimator are also shown.  Notice that the marginal likelihood values computed by the learnt harmonic mean estimator are highly accurate and are indeed sensitive to changes in the prior. }
  \label{fig:normal_gamma_comparison}
\end{figure}

The marginal likelihood values computed analytically and using our learnt harmonic mean estimator are shown in \tbl{\ref{tbl:normal_gamma_comparison}} for different priors as $\tau_0$ is varied.  For comparison, the errors between the analytic values and those estimated by our learnt harmonic mean estimator and the original harmonic mean estimator are also shown.  Notice that the marginal likelihood values computed by the learnt harmonic mean estimator are highly accurate and do indeed vary with differing priors, in contrast to results computed by the original harmonic mean estimator \citep{friel:2012}.  The improvement in accuracy between the original and our learnt harmonic mean estimation is approximately four orders of magnitude in log space.  In addition, to graphically compare the marginal likelihood values estimated by the learnt harmonic mean estimator to the analytic values, we plot in \fig{\ref{fig:normal_gamma_comparison}} the ratio of the estimated and analytic values, with the uncertainties computed by our learnt harmonic mean estimator overlaid.  It is clear that the marginal likelihood values computed by the learnt harmonic mean estimator closely estimate the analytic values and that the uncertainties are reasonable.

%-------------------------------------------------------------------------------
\subsection{Logistic regression models: Pima Indian example}
%-------------------------------------------------------------------------------

We consider the comparison of two logistic regression models using the \textit{Pima Indians} data, which is another common benchmark problem for comparing estimators of the marginal likelihood.  The original harmonic mean estimator has been shown to fail catastrophically for this example \citep{friel:2012}, whereas we show here that our learnt harmonic mean estimator is highly accurate.

The Pima Indians data \citep{smith:1988}, originally from the National Institute of Diabetes and Digestive and Kidney Diseases, were compiled from a study of indicators of diabetes in $n=532$ Pima Indian women aged 21 or over.  Seven primary predictors of diabetes were recorded, including: number of prior pregnancies (NP);  plasma glucose concentration (PGC); diastolic blood pressure (BP); triceps skin fold thickness (TST); body mass index (BMI); diabetes pedigree function (DP); and age (AGE).

\begin{figure}
  \begin{center}
    % !TeX root = harmonic_cosmo21_2018.tex

\begin{tikzpicture}[scale=1.0, transform shape]
  [squarednode/.style={rectangle, draw=black, minimum size=8mm},
    latent/.style={circle, draw=black, minimum size=10mm}]

  % \footnotesize

  % Covariate 1 -->
  \node[obs, minimum size=0.95cm] (X1) {\small NP$_i$};

  % Covariate 2 -->
  \node[obs, right=1cm of X1, minimum size=0.95cm] (X2) {\small PGC$_i$};

  % Covariate 3 -->
  \node[obs, right=1cm of X2,  minimum size=0.95cm] (X3) {\small BMI$_i$};

  % Covariate 4 -->
  \node[obs, right=1cm of X3, minimum size=0.95cm] (X4) {\small DP$_i$};

  % Covariate 5 -->
  \node[obs, right=1cm of X4,  minimum size=0.95cm] (X5) {\small AGE$_i$};

  % Define nodes
  \node[latent, below=.5cm of X2, xshift=1cm]  (P) {\large $p_{i}$};
  \node[obs,below=.5cm of P] (y) {\large $y_i$};

  % Theta --> P
  \node[latent, above=3.25cm of y, xshift=0.0cm]  (theta) {\large $\theta$};
  \node[factor, above=0.5cm of theta, xshift=0cm]  (N) {N};
  \node[const, right=0.1cm of N] () {\normalsize Normal};
  \node[latent, above=0.5cm of N, xshift=0.0cm]  (tau) {\large $\tau$};

  \draw [->] (P) -- (y);

  \draw [->] (X1) -- (P);
  \draw [->] (X2) -- (P);
  \draw [->] (X3) -- (P);
  \draw [->] (X4) -- (P);
  \draw [->] (X5) -- (P);

  \draw [->] (theta) -- (P);

  \factoredge{tau}{N}{theta}

  \platenode {} {(X1)(X1)(P)(X5)(y)} {$i \in \lbrace 1, ..., n \rbrace$}; %

\end{tikzpicture}
    \caption{Graphical representation of logistic regression Model 2 for modelling diabetes in Pima Indians.  Model 1 is similar but does not include the AGE covariate.}
    \label{fig:hbm_pima_indians}
  \end{center}
\end{figure}
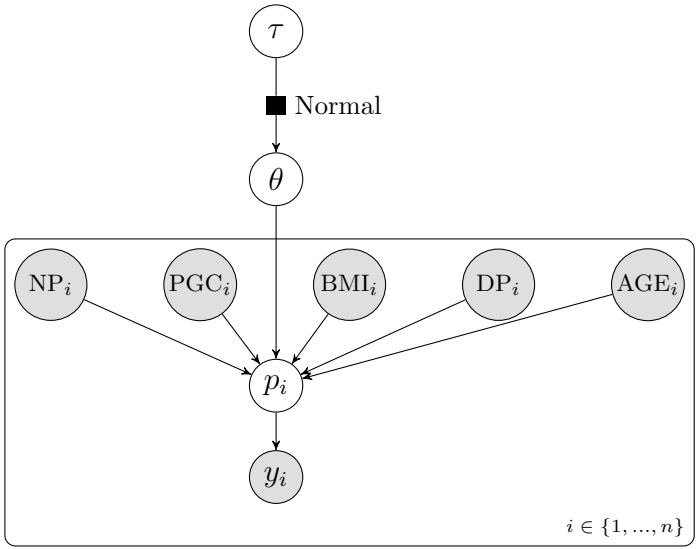

The probability of diabetes $p_i$ for person $i \in \{1, \ldots, n\}$ can be modelled by the standard logistic function
\begin{equation}
  p_i = \frac{1}{1+\exp\bigl(- \theta^\text{T} x_i\bigr)}
  \spcend ,
\end{equation}
with covariates $x_i = (1,x_{i,1}, \dots x_{i,d})^\text{T}$ and parameters $\theta = (\theta_0, \dots, \theta_d)^\text{T}$, where $d$ is the total number of covariates considered.  The likelihood function then reads
\begin{equation}
  \mathcal{L}({y} \given {\theta}) = \prod_{i=1}^n p_i^{y_i}(1-p_i)^{1-y_i}
  \spcend ,
\end{equation}
where $y = (y_1, \dots, y_n)^\text{T}$ is the diabetes incidence, \textit{i.e.} $y_i$ is unity if patient $i$ has diabetes and zero otherwise.
An independent multivariate Gaussian prior is assumed for the parameters $\theta$, given by
\begin{equation}
  \pi(\theta)
  = \Bigl(  \frac{\tau}{2\pi} \Bigr)^{d/2}
  \exp \bigl( - \frac{\tau}{2} \theta^\text{T} \theta \bigr)
  \spcend ,
\end{equation}
with precision $\tau$.

Two different logistic regression models are considered, with different subsets of covariates:
\begin{align*}
  \text{Model } M_1: & \quad \text{covariates = \{NP, PGC, BMI, DP\} (and bias);}      \\*
  \text{Model } M_2: & \quad \text{covariates = \{NP, PGC, BMI, DP, AGE\} (and bias).}
\end{align*}
A graphical representation of Model 2 is illustrated in \fig{\ref{fig:hbm_pima_indians}} (Model 1 is similar but does not include the AGE covariate).

We compute the marginal likelihood for both Model 1 and Model 2 using our learnt harmonic mean estimator for $\tau = 0.01$ and $\tau = 1$, as in \cite{friel:2012}.  A reversible jump algorithm \citep{green:1995} is used by \citet{friel:2012} to compute benchmark Bayes factors $\text{BF}_{12}$ of 13.96 and 1.30, respectively, for $\tau = 0.01$ and $\tau = 1$, which are treated as ground truth values.

For our learnt harmonic mean estimator we use \texttt{emcee} to draw 5,000 samples for 200 chains, with burn in of 1,000 samples, yielding 4,000 posterior samples per chain.  We use 25\% of the samples to learn the target model, using cross-validation to select between the hypersphere and modified Gaussian mixture model.  In all cases the modified Gaussian mixture model is selected.  The remaining 75\% of posterior samples are used for inferring the marginal likelihood.  Computation time is typically a few minutes on a standard laptop, including drawing samples (note that fewer samples could likely be used to reduce computation time if required).

The marginal likelihood values computed by our learnt harmonic mean estimator are shown in \tbl{\ref{tbl:pima_tau_0p01}} and \tbl{\ref{tbl:pima_tau_1p00}} for the cases $\tau = 0.01$ and $\tau = 1$, respectively.  The benchmark values and errors of both the standard and learnt harmonic mean estimator are also shown for comparison.  While the standard harmonic mean estimator fails catastrophically on this problem \citep{friel:2012}, our learnt harmonic mean estimator is robust and highly accurate.

\begin{table}
  \caption{Marginal likelihood values computed by the learnt harmonic mean estimator for the Pima Indians logistic regression models for prior precision $\tau = 0.01$. While the original harmonic mean estimator fails catastrophically, our learnt harmonic mean estimator is highly accurate.}% ${}^*$Denotes computed by \cite{friel:2012}.}
  \label{tbl:pima_tau_0p01}
  \centering
  \begin{tabular}{lrrr}\toprule
              & \multicolumn{1}{c}{Model $M_1$} & \multicolumn{1}{c}{Model $M_2$} & \multicolumn{1}{c}{$\log\text{BF}_{12}$}      \\
              & \multicolumn{1}{c}{$\log(z_1)$} & \multicolumn{1}{c}{$\log(z_2)$} & \multicolumn{1}{c}{$= \log(z_1) - \log(z_2)$} \\ \midrule
    Benchmark & --                              & --                              & 2.63620                                       \\
    Estimated & -257.23656                      & -259.86669                      & 2.63014                                       \\
              & $\pm\,$0.00264                  & $\pm\,$0.00968                  & $\pm\,$0.01232                                \\
    Error     & --                              & --                              & 0.00606                                       \\
    {\scriptsize (learnt harmonic mean)}\hspace*{0mm}                                                                             \\ \midrule
    Error     & --                              & --                              & -2.67760                                      \\
    {\scriptsize (original harmonic mean)}\hspace*{0mm}                                                                           \\ \bottomrule
  \end{tabular}
\end{table}

\begin{table}
  \caption{Marginal likelihood values computed by the learnt harmonic mean estimator for the Pima Indians logistic regression models for prior precision $\tau = 1.0$. While the original harmonic mean estimator fails catastrophically, our learnt harmonic mean estimator is highly accurate.}% ${}^*$Denotes computed by \cite{friel:2012}.}
  \label{tbl:pima_tau_1p00}
  \centering
  \begin{tabular}{lrrr}\toprule
              & \multicolumn{1}{c}{Model $M_1$} & \multicolumn{1}{c}{Model $M_2$} & \multicolumn{1}{c}{$\log\text{BF}_{12}$}      \\
              & \multicolumn{1}{c}{$\log(z_1)$} & \multicolumn{1}{c}{$\log(z_2)$} & \multicolumn{1}{c}{$= \log(z_1) - \log(z_2)$} \\ \midrule
    Benchmark & --                              & --                              & 0.26236                                       \\
    Estimated & -247.30633                      & -247.56128                      & 0.25495                                       \\
              & $\pm\,$0.00239                  & $\pm\,$0.00789                  & $\pm\,$0.01028                                \\
    Error     & --                              & --                              & 0.00742                                       \\
    {\scriptsize (learnt harmonic mean)}\hspace*{0mm}                                                                             \\ \midrule
    Error     & --                              & --                              & -0.44567                                      \\
    {\scriptsize (original harmonic mean)}\hspace*{0mm}                                                                           \\ \bottomrule
  \end{tabular}
\end{table}

%-------------------------------------------------------------------------------
\subsection{Non-nested linear regression models: Radiata pine example}
%-------------------------------------------------------------------------------

We consider another example where the original harmonic mean estimator was shown to fail catastrophically \citep{friel:2012}.  In particular, we consider non-nested linear regression models for the \textit{Radiata pine} data, which is another common benchmark data-set \citep{williams:1959}, and show that our learnt harmonic mean estimator is highly accurate.

For $n=42$ trees, the Radiata pine data-set includes measurements of the maximum compression strength parallel to the grain $y_i$, density $x_i$ and resin-adjusted density $z_i$, for specimen $i \in \{1, \ldots, n\}$.  The question at hand is whether density or resin-adjusted density is a better predictor of compression strength.  This motivates two Gaussian linear regression models:
\begin{align}
  \text{Model } M_1 & :          & \datum_i                             & = \alpha + \beta(x_i - \bar{x}) + \epsilon_i ,
                    & \epsilon_i & \sim \text{N}(0, \tau^{-1}) \spcend;                                                  \\
  \text{Model } M_2 & :          & \datum_i                             & = \gamma + \delta(z_i - \bar{z}) + \eta_i ,
                    & \eta_i     & \sim \text{N}(0, \lambda^{-1})
  \spcend,
\end{align}
where $\bar{x} = \frac{1}{n} \sum_{i=1}^n x_i$, $\bar{z} = \frac{1}{n} \sum_{i=1}^n z_i$, and $\tau$ and $\lambda$ denote the precision (inverse variance) of the noise for the respective models.

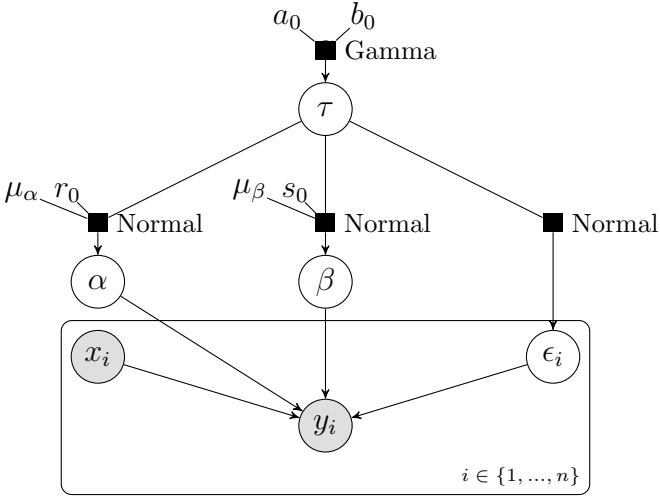
\begin{figure}
  \begin{center}
    % !TeX root = harmonic_cosmo21_2018.tex

\begin{tikzpicture}[scale=1.0, transform shape]
  [squarednode/.style={rectangle, draw=black, minimum size=8mm},
    latent/.style={circle, draw=black, minimum size=10mm}]

  % \footnotesize

  % Define nodes
  \node[obs] (y) {\large $y_i$};

  \node[obs, above=0.2cm of y, xshift=-3.0cm] (x) {\large $x_i$};

  \node[latent, above=0.2cm of y, xshift=3.0cm]  (epsilon) {\large $\epsilon_i$};

  \node[latent, above=3.5cm of y, xshift=0cm]  (tau) {\large $\tau$};
  \node[factor, above=0.3cm of tau, xshift=0cm]  (G1) {G};
  \node[const, right=0.1cm of G1] () {\normalsize Gamma};
  \node[const, above=0.7cm of tau, xshift=-0.5cm] (a0) {\large $a_0$};
  \node[const, above=0.7cm of tau, xshift=0.5cm] (b0) {\large $b_0$};

  \node[latent, above=1.2cm of y, xshift=-3.0cm]  (alpha) {\large $\alpha$};
  \node[factor, above=0.3cm of alpha, xshift=0cm]  (N2) {N};
  \node[const, right=0.1cm of N2] () {\normalsize Normal};
  \node[const, above=0.7cm of alpha, xshift=-1.0cm] (mu_alpha) {\large $\mu_\alpha$};
  \node[const, above=0.7cm of alpha, xshift=-0.4cm] (r0) {\large $r_0$};

  \node[latent, above=1.2cm of y, xshift=0.0cm]  (beta) {\large $\beta$};
  \node[factor, above=0.3cm of beta, xshift=0cm]  (N3) {N};
  \node[const, right=0.1cm of N3] () {\normalsize Normal};
  \node[const, above=0.7cm of beta, xshift=-1.0cm] (mu_beta) {\large $\mu_\beta$};
  \node[const, above=0.7cm of beta, xshift=-0.4cm] (s0) {\large $s_0$};

  \node[factor, above=0.3cm of beta, xshift=3.0cm]  (N1) {N};
  \node[const, right=0.1cm of N1] () {\normalsize Normal};

  \draw [->] (x) -- (y);
  \draw [->] (alpha) -- (y);
  \draw [->] (beta) -- (y);
  \draw [->] (epsilon) -- (y);

  \factoredge{tau}{N1}{epsilon};
  \factoredge{tau,mu_alpha,r0}{N2}{alpha};
  \factoredge{tau,mu_beta,s0}{N3}{beta};
  \factoredge{a0,b0}{G1}{tau};

  \platenode {} {(y)(x)(epsilon)} {$i \in \lbrace 1, ..., n \rbrace$}; %

\end{tikzpicture}
    \caption{Graphical representation of the non-nested linear regression Model 1 for modelling maximum compression strength for Radiata pine.  Model 2 is similar.}
    \label{fig:hbm_radiata_pine}
  \end{center}
\end{figure}

For Model 1, Gaussian priors are assumed for the bias and linear terms:
\begin{align}
  \alpha & \sim \text{N}\bigl(\mu_\alpha, (r_0 \tau)^{-1}\bigr) \spcend ; \\
  \beta  & \sim \text{N}\bigl(\mu_\beta, (s_0 \tau)^{-1}\bigr) \spcend ,
\end{align}
with means $\mu_\alpha = 3000$ and $\mu_\beta = 185$, and precision scales $r_0 = 0.06$ and $s_0 = 6$.  A gamma prior is assumed for the noise precision:
\begin{equation}
  \tau \sim \text{Ga}(a_0, b_0) \spcend ,
\end{equation}
with shape $a_0 = 3$ and rate $b_0 = 2 \times 300^2$.  The joint prior for $(\alpha, \beta, \tau)$ then reads:
\begin{align}
  \pi(\alpha, \beta, \tau)
   & = \pi(\alpha, \beta \given \tau) \pi(\tau)                     \\*
   & = \pi(\alpha \given \tau) \pi(\beta \given \tau) \pi(\tau)     \\*
   & = \frac{(b_0\tau_0)^{a_0} (r_0 s_0)^{1/2} }{2 \pi \Gamma(a_0)}
  \exp\bigl(-b_0 \tau\bigr) \nonumber                               \\*
   & \quad\quad \times
  \exp\Bigl(-\frac{\tau}{2}\bigl(r_0(\alpha-\mu_\alpha)^2 + s_0(\beta-\mu_\beta)^2\bigr)\Bigr)
  \spcend .
\end{align}
The likelihood for Model 1 is given by
\begin{align}
  \mathcal{L}({x}, {y})
   & = \prod_{i=1}^n \text{P}(x_i, y_i \given \alpha, \beta, \tau)                    \\*
   & = \prod_{i=1}^n \sqrt{\frac{\tau}{2\pi}}
  \exp\Bigl(- \frac{\tau}{2} \bigl(y_i - \alpha - \beta (x_i - \bar{x})\bigr)^2\Bigr) \\
   & = \Bigl(\frac{\tau}{2\pi}\Bigr)^{n/2}
  \exp\Bigl(- \frac{\tau}{2} \sum_{i=1}^n \bigl(y_i - \alpha - \beta (x_i - \bar{x})\bigr)^2\Bigr)
  \spcend ,
\end{align}
where $x = (x_1, \dots, x_n)^\text{T}$ and $y = (y_1, \dots, y_n)^\text{T}$.  For Model 2, the priors adopted for $(\gamma, \delta, \lambda)$ are the same as those adopted for $(\alpha, \beta, \tau)$ of Model 1, respectively, with the same hyperparameters.  The likelihood for Model 2 again takes an identical form to Model 1.  A graphical representation of Model 1 is illustrated in \fig{\ref{fig:hbm_radiata_pine}} (Model 2 is similar).

One reason this problem has become a common benchmark for comparing marginal likelihood estimators is that the marginal likelihood of the two models can be computed analytically.  The evidence for Model 1 is given by
\begin{equation}
  \label{eqn:evidence_radiata_pine}
  z = \frac{(2b_0)^{a_0}}{\pi^{n/2}} \frac{\Gamma(a_0 + n/2)}{\Gamma(a_0)}
  \frac{\vert Q_0 \vert^{1/2}}{\vert M\vert^{1/2}}
  \bigl({y}^\text{T}{y}
  + {\mu_0}^\text{T} Q_0 {\mu_0}
  - {\nu_0}^\text{T} M {\nu_0} + 2 b_0\bigr)^{-a_0-n/2}
  \spcend ,
\end{equation}
where ${\mu_0} = (\mu_\alpha, \mu_\beta)^\text{T}$, $Q_0 = \text{diag}(r_0, s_0)$, $M = X^\text{T} X + Q_0$, and ${\nu_0} = M^{-1} (X^\text{T} {y} + Q_0 {\mu_0})$, for the feature matrix $X$ with row $i$ containing $(1,\  x_i-\bar{x})$.

We compute the marginal likelihood analytically using \eqn{\ref{eqn:evidence_radiata_pine}} and also numerically using our learnt harmonic mean estimator.  For the learnt harmonic mean estimator we use \texttt{emcee} to draw 20,000 samples for 400 chains, with burn in of 2,000 samples.  We use 25\% of the samples to learn the target model, adopting a simple hypersphere model, and use the remaining 75\% for inferring the marginal likelihood.  Computation time is a few minutes on a standard laptop, including drawing samples (note that fewer samples could likely be used to reduce computation time if required).
% (note that accurate results can likely be estimated with many fewer samples but since computation time is already low we do not focus on optimising the number of samples).

The analytic and estimated marginal likelihood values are shown in \tbl{\ref{tbl:radiata_pine}} for the two models, with the Bayes factor comparing the two models.  For the values computed by our learnt harmonic mean estimator we also show the estimated uncertainty.  The errors of both the standard and learnt harmonic mean estimator are shown for comparison. Note that the uncertainties estimated by the learnt harmonic mean estimator appear reasonable.  While the standard harmonic mean estimator fails catastrophically on this problem \citep{friel:2012}, our learnt harmonic mean estimator is highly accurate.

\begin{table}
  \caption{Marginal likelihood values computed analytically and by the learnt harmonic mean estimator for the Radiata pine non-nested linear regression models.   While the original harmonic mean estimator fails catastrophically, our learnt harmonic mean estimator is highly accurate.}% ${}^*$Denotes computed by \cite{friel:2012}.}
  \label{tbl:radiata_pine}
  \centering
  \begin{tabular}{lrrr}\toprule
              & \multicolumn{1}{c}{Model $M_1$} & \multicolumn{1}{c}{Model $M_2$} & \multicolumn{1}{c}{$\log\text{BF}_{21}$}      \\
              & \multicolumn{1}{c}{$\log(z_1)$} & \multicolumn{1}{c}{$\log(z_2)$} & \multicolumn{1}{c}{$= \log(z_2) - \log(z_1)$} \\ \midrule
    Analytic  & -310.12829                      & -301.70460                      & 8.42368                                       \\
    Estimated & -310.12807                      & -301.70413                      & 8.42394                                       \\
              & $\pm\,$0.00072                  & $\pm\,$0.00074                  & $\pm\,$0.00145                                \\
    Error     & 0.00022                         & 0.00047                         & 0.00026                                       \\
    {\scriptsize (learnt harmonic mean)}\hspace*{0mm}                                                                             \\ \midrule
    Error     & --                              & --                              & -0.17372                                      \\
    {\scriptsize (original harmonic mean)}\hspace*{0mm}                                                                           \\ \bottomrule
  \end{tabular}
\end{table}

%-------------------------------------------------------------------------------
\subsection{Gaussian in varying dimensions}
%-------------------------------------------------------------------------------

Finally, we illustrate the application of our estimator beyond low-dimensional settings, considering experiments where the dimension of the parameter space increases.  For simplicity we consider a Gaussian likelihood with a uniform prior, where the marginal likelihood can be computed analytically.  We adopt the simple hypersphere model, which is effective for a Gaussian posterior.  Results are illustrated in \tbl{\ref{tbl:gaussian}}.  Note that parameters were not optimised and accurate results could likely be obtained with fewer samples. Further note that the computation times recorded do not include the time accumulated during initial burn-in.  It is apparent that the estimator is accurate in settings with dimensions $\order(10^3)$ and potentially beyond.

\begin{table}
  \caption{Marginal likelihood values computed analytically and by the learnt harmonic mean estimator for a Gaussian posterior in varying dimensions.  Note that parameters were not optimised and results could likely be computed to comparable accuracy with fewer samples (i.e.\ with lower computation time).  Further note that the computation times recorded do not include the time accumulated during initial burn-in.
  }
  \label{tbl:gaussian}
  \centering
  \begin{tabular}{crrrr}\toprule
    Dimension & \multicolumn{1}{c}{Analytic}  & \multicolumn{1}{c}{Estimated} & \multicolumn{1}{c}{Error} & \multicolumn{1}{c}{Computation} \\
              & \multicolumn{1}{c}{$\log(z)$} & \multicolumn{1}{c}{$\log(z)$} & \multicolumn{1}{c}{(\%)}  & \multicolumn{1}{c}{time}        \\ \midrule
    32        & -29.406                       & -29.411                       & 0.0180\%                  & $\sim$10 sec                    \\
    64        & -58.812                       & -58.813                       & 0.0008\%                  & $\sim$20 sec                    \\
    128       & -117.62                       & -117.63                       & 0.0026\%                  & $\sim$3 min                     \\
    256       & -235.25                       & -235.25                       & 0.0015\%                  & $\sim$18 min                    \\
    512       & -470.50                       & -470.49                       & 0.0006\%                  & $\sim$20 min                    \\
    1024      & -940.99                       & -941.06                       & 0.0073\%                  & $\sim$3 hours                   \\\bottomrule
  \end{tabular}
  % \begin{tabular}{cccccccc}\toprule
  %   Dimension & Analytic  & Estimated & Error    & Number & Samples       & Burn-in         & Computation      \\
  %             & $\log(z)$ & $\log(z)$ & (\%)     & chains & per chain     &                 & time             \\ \midrule
  %   32        & -29.406   & -29.411   & 0.0180\% & 64     & $10^4$        & $7\times10^3$   & $\sim$9 sec \\
  %   64        & -58.812   & -58.813   & 0.0008\% & 128    & $10^4$        & $7\times10^3$   & $\sim$18 sec  \\
  %   128       & -117.62   & -117.63   & 0.0026\% & 256    & $10^4$        & $7\times10^3$   & $\sim$3 min \\
  %   256       & -235.25   & -235.25   & 0.0015\% & 512    & $10^4$        & $7\times10^3$   & $\sim$18 min \\
  %   512       & -470.50   & -470.49   & 0.0006\% & 1024   & $5\times10^4$ & $4.7\times10^4$ & $\sim$20 min \\
  %   1024      & -940.99   & -941.06   & 0.0073\% & 2048   & $10^5$        & $8.7\times10^4$ & $\sim$3 hours  \\\bottomrule
  % \end{tabular}
\end{table}

%-------------------------------
\subsection{Cosmological example}
%-------------------------------

% intro
We conclude our numerical experiments with an application of the learnt harmonic mean estimator to a model comparison problem in cosmology. We consider measurements of the Cosmic Microwave Background (CMB) temperature ($T$) and polarisation ($E$) anisotropies from Data Release 4 of the Atacama Cosmology Telescope (ACT) \citep{Aiola_2020, Choi_2020}.

% cosmological models and parameters
The dataset consists of CMB power spectra $C_{\ell}^{XY}$, with $\{X, Y\} = \{T, E\}$ measured at angular multipoles $\ell$ up to 5000. CMB power spectra are functions of the cosmological parameters that describe the model assumed for the composition and evolution of the Universe. The $\Lambda$ Cold Dark Matter ($\Lambda$CDM) model is currently regarded as the concordance cosmological scenario, as it provides an excellent fit to numerous cosmological measurements. However, some of its key components are poorly understood, including the cosmological constant $\Lambda$ that gives its name to the model and is deemed responsible for the accelerated expansion of the Universe. Alternative models have been proposed to account for cosmic acceleration. One of them attributes this phenomenon to a dynamic dark energy component -- a fluid whose equation of state (the relation between its pressure $P$ and density $\rho$) changes over time. Using redshift $z$ as a proxy for cosmic time (with $z=0$ today), this can be written as $P=w(z)\rho$. A common parameterisation for $w(z)$ is $w(z) = w_0 + w_a\frac{z}{1+z}$ \citep{Chevallier_2001, Linder_2003}, with $w_0 = -1$ and $w_a=0$ corresponding to a cosmological constant. Thus, the $w_0 w_a$CDM model can be seen as an extension to the $\Lambda$CDM model, to account for cosmic acceleration through a dynamic dark energy field which introduces two additional cosmological parameters in the model, $w_0$ and $w_a$.

% general method
We aim to perform a comparison between the $\Lambda$CDM and $w_0 w_a$CDM models using ACT CMB data. We employ the learnt harmonic mean estimator to derive estimates of the Bayesian evidence for each model, and the Bayes factor between the two. We then compare these numbers against those obtained with an independent approach, namely a nested sampler which produces an estimate of the evidence as well as posterior samples for each model. To run our experiments we use the publicly available \texttt{pyactlike} likelihood \footnote{\href{https://github.com/ACTCollaboration/pyactlike/}{https://github.com/ACTCollaboration/pyactlike/}}, developed by the ACT collaboration to analyse their CMB power spectra measurements. This likelihood only includes one additional `nuisance' parameter, $y_p^2$, to the cosmological parameters underlying the model of the Universe. This parameter acts as a phenomenological rescaling of the spectra to account for undetected systematics. Table~\ref{tab:priors_cosmo} shows the priors assumed for all of the parameters varied in the analysis. Priors on all parameters are shared between the \texttt{emcee} and \texttt{PolyChord} runs, with the exception of $w_0$ and $w_a$ used only in the dynamic dark energy scenario.

% PolyChord vs emcee setups
We compare two experimental setups:
\begin{enumerate}
  \item In the first one, we use the affine sampler \texttt{emcee} \citep{Foreman_Mackey_2013} to produce 30,000 samples of the posterior, using 300 walkers each collecting 150 samples, of which we discard 50 for each walker as burn-in. We run the sampler on a 60-core parallelised configuration, which takes approximately 5 hours to complete. Once obtained the posterior samples, we use our implementation of the learnt harmonic mean estimator to derive an estimate of the evidence for each of the two cosmological models. We use the hypersphere model for the learnt harmonic mean estimator in both cosmological scenarios, and perform a 50:50 split of the posterior samples for the training and testing phase.

  \item In the second setup, we run the nested sampler \texttt{PolyChord} \citep{Handley_2015a, Handley_2015b} to obtain posterior samples and an estimate of the evidence for each model. We run \textsc{PolyChord} through the \texttt{Cobaya} \citep{Torrado_2021, Torrado_2021_software} software for cosmological data analysis. We use the same parallelised configuration over 60 cores used for the \texttt{emcee} runs, and default values for the \texttt{PolyChord} options. This leads to $\sim 5,000$ and $10,000$ samples for the $\Lambda$CDM and $w_0 w_a$CDM model, respectively, obtained in $\sim$15 hours for both cases.
\end{enumerate}
\begin{table}
  \caption{Priors for the parameters varied in the cosmological analyses. Note that all of the parameters are shared between the $\Lambda$CDM and $w_0 w_a$CDM model except for $w_0$ and $w_a$. $\mathcal{U}$ denotes a Uniform distribution, while $\mathcal{N}$ a Gaussian one. All priors are kept the same between the \texttt{emcee} and \texttt{PolyChord} runs.}
  \centering
  \begin{tabular}{c c}\toprule
    Parameter         & Prior                        \\ \midrule
    $\omega_b$        & $\mathcal{U}(0.005, 0.01)$   \\
    $\omega_{cdm}$    & $\mathcal{U}(0.001,  0.99)$  \\
    $100 \, \theta_s$ & $\mathcal{U}(0.5,    10.)$   \\
    $\tau$            & $\mathcal{N}(0.065,  0.015)$ \\
    $n_s$             & $\mathcal{U}(0.8,    1.2)$   \\
    $\ln 10^{10}A_s$  & $\mathcal{U}(1.61,   3.91)$  \\
    $y_p^2$           & $\mathcal{U}(0.9,    1.1)$   \\
    \midrule
    $w_0$             & $\mathcal{U}(-1.5,   -0.5)$  \\
    $w_a$             & $\mathcal{U}(-0.5,   0.5)$   \\ \bottomrule
  \end{tabular}
  \label{tab:priors_cosmo}
\end{table}
\begin{figure}
  \centering
  \includegraphics[width=\textwidth]{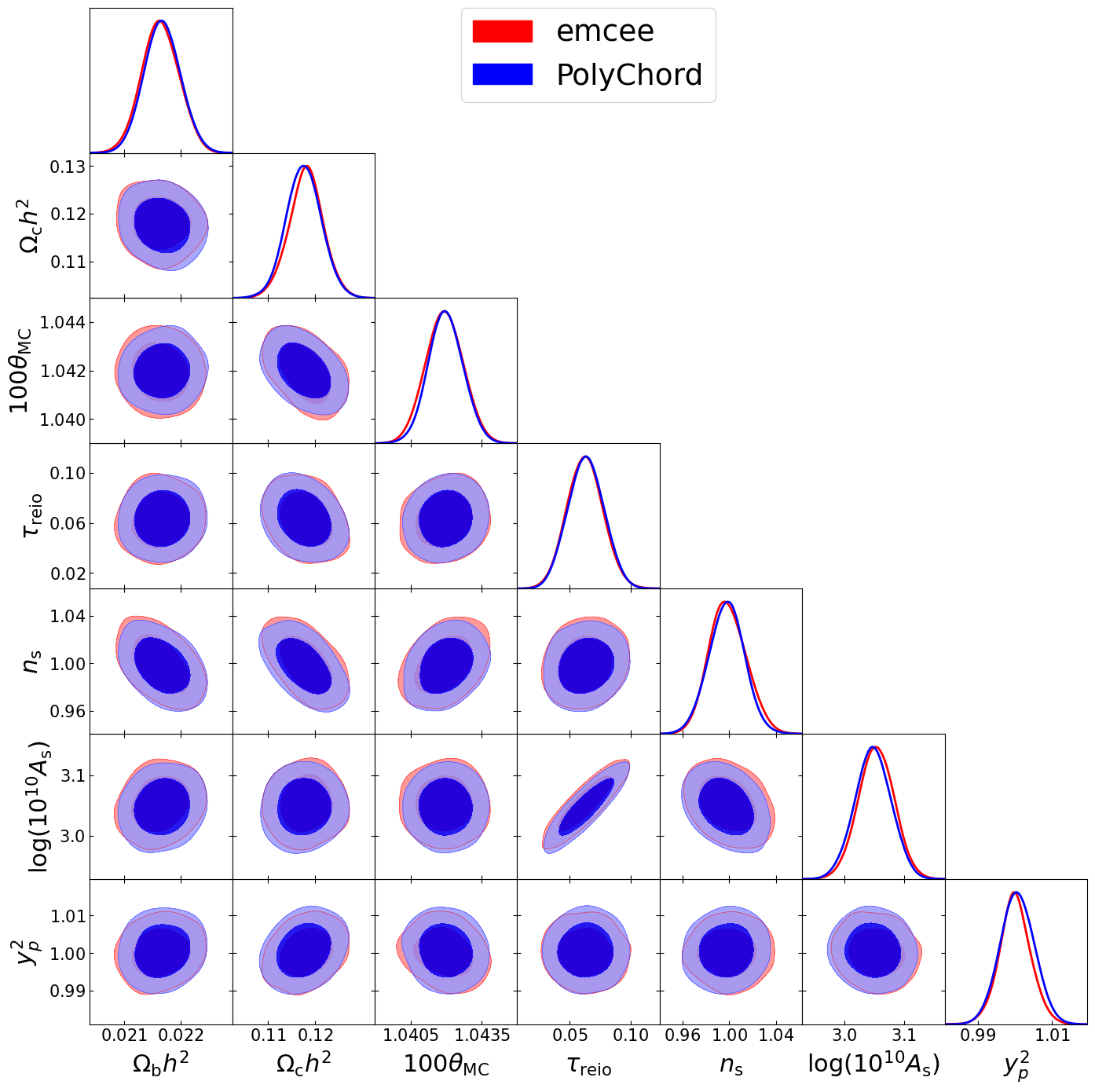}
  \caption{Marginalised 68 \% and 95\% 2D posterior contours and 1D marginalised posterior distribution for the parameters varied in the $\Lambda$CDM runs. \textit{red} contours have been obtained using \texttt{emcee}, while \textit{blue} contours with \texttt{PolyChord}.}
  \label{fig:contours_cosmo_lcdm}
\end{figure}
\begin{figure}
  \centering
  \includegraphics[width=\textwidth]{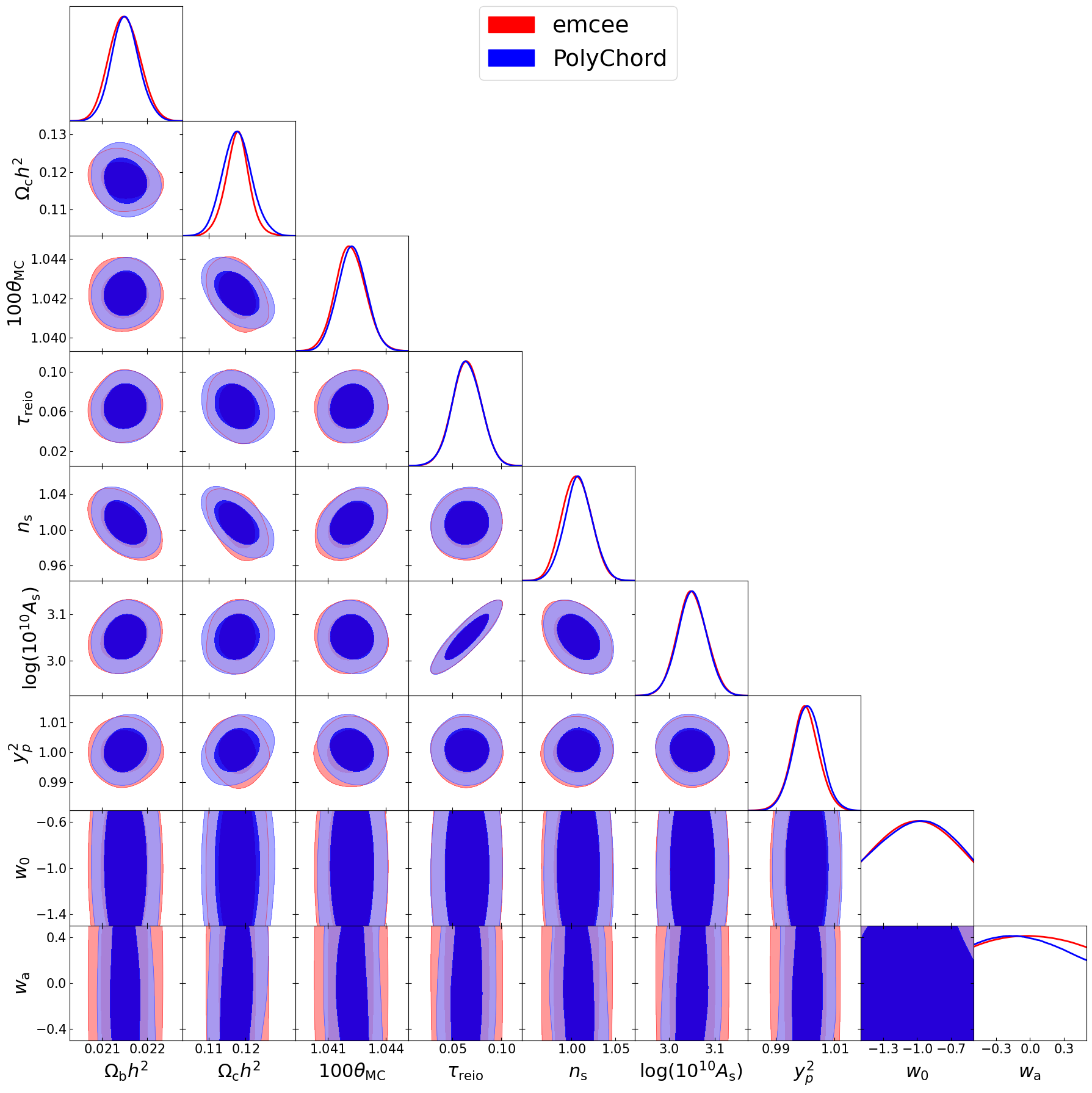}
  \caption{Same as Fig.~\ref{fig:contours_cosmo_lcdm}, but for the $w_0 w_a$CDM model.}
  \label{fig:contours_cosmo_w0wa}
\end{figure}

% results
Fig.~\ref{fig:contours_cosmo_lcdm} and \ref{fig:contours_cosmo_w0wa} show the posterior contours for the $\Lambda$CDM and $w_0 w_a$CDM model, respectively. Each figure compares contours obtained with the \texttt{emcee} (\textit{red}) and \texttt{PolyChord} (\textit{blue}) sampler. For the $\Lambda$CDM runs, we obtain a value of the log-evidence $\log Z$
\begin{align}
  \log Z & = -168.92 \pm 0.35 \, (\textrm{\texttt{PolyChord}}) \nonumber   \\
  \log Z & = -168.87 \pm 0.29 \, (\textrm{learnt harmonic mean estimator})
\end{align}
while for $w_0 w_a$CDM we obtain
\begin{align}
  \log Z & = -169.38 \pm 0.24 \, (\textrm{\texttt{PolyChord}}), \nonumber   \\
  \log Z & = -169.32 \pm 0.25 \, (\textrm{learnt harmonic mean estimator}).
\end{align}
Combined, these measurements lead to a Bayes factor $\Delta \log Z = \log Z_{\Lambda \rm{CDM}} - \log Z_{w_0 w_a \rm{CDM}}$
\begin{align}
  \Delta \log Z & = 0.46 \pm 0.42 \, (\textrm{\texttt{PolyChord}}), \nonumber   \\
  \Delta \log Z & = 0.45 \pm 0.38 \, (\textrm{learnt harmonic mean estimator}).
\end{align}
We can see from these results that the learnt harmonic mean estimator leads to numerical conclusions on the comparison between the $\Lambda$CDM and $w_0 w_a$CDM models in excellent agreement with those derived from running \texttt{PolyChord}, \textit{i.e.} mildly favouring $\Lambda$CDM. Note also that by virtue of its independence from the sampling method, the learnt harmonic mean estimator allows the user to resort to the \texttt{emcee} sampler which leads to $\sim$ 3 to 6 times more samples than \texttt{PolyChord} in 1/3 of the time, using the same hardware resources.

%===============================================================================
\section{Software package}
\label{sec:code}
%===============================================================================

\definecolor{codegreen}{rgb}{0,0.6,0}
\definecolor{codegray}{rgb}{0.5,0.5,0.5}
\definecolor{codepurple}{rgb}{0.58,0,0.82}
\definecolor{backcolour}{rgb}{0.95,0.95,0.92}

\lstdefinestyle{mystyle}{
  backgroundcolor=\color{backcolour},
  commentstyle=\color{codegreen},
  keywordstyle=\color{magenta},
  numberstyle=\tiny\color{codegray},
  stringstyle=\color{codepurple},
  basicstyle=\footnotesize,
  breakatwhitespace=false,
  breaklines=true,
  captionpos=b,
  keepspaces=true,
  numbers=left,
  numbersep=5pt,
  showspaces=false,
  showstringspaces=false,
  showtabs=false,
  tabsize=2
}

\lstset{style=mystyle}

The learnt harmonic mean estimator is implemented in the \texttt{harmonic} software package\footnote{\url{https://github.com/astro-informatics/harmonic}}, which is open source and publicly available.  Careful consideration has been given to the design and implementation of the code, following software engineer best practices (for example, at the time of release test coverage is over 96\%).

Since the learnt harmonic mean estimator requires samples from the posterior distribution only, the \texttt{harmonic} code is agnostic to the method or code used to generate posterior samples.  That said, \texttt{harmonic} works exceptionally well with MCMC sampling techniques that naturally provide samples from multiple chains by their ensemble nature, such as affine invariance ensemble samplers \citep{goodman:2010}.  As discussed in \sectn{\ref{sec:review}}, we advocate running a number of independent MCMC chains and using all of the correlated samples within a chain to avoid the loss of efficiency that otherwise results from thinning an MCMC chain.  The \texttt{emcee} code\footnote{\url{https://emcee.readthedocs.io/en/stable/}} \citep{foreman-mackey:2013} provides an excellent implementation of the affine invariance ensemble samplers proposed by \citet{goodman:2010}.  \texttt{emcee} is thus a natural choice for use with \texttt{harmonic} and we have specifically designed \texttt{harmonic} to ensure it works seamlessly with \texttt{emcee} (although of course other samplers can also be considered).  In code Listing~1 we give an example of usage of \texttt{harmonic} with \texttt{emcee} to demonstrate how easy it is to use the combination for marginal likelihood estimation.

\pagebreak

\begin{lstlisting}[language=Python, caption={Example usage of \texttt{harmonic} to compute the marginal likelihood, using \texttt{emcee} to perform MCMC sampling.}]
import numpy as np
import emcee
import harmonic

# Run sampler
sampler = emcee.EnsembleSampler(nchains, ndim, ln_posterior,
                                args=[posterior_args])
(pos, prob, state) = sampler.run_mcmc(pos, samples_per_chain)
samples = np.ascontiguousarray(sampler.chain[:,nburn:,:])
lnprob = np.ascontiguousarray(sampler.lnprobability[:,nburn:])

# Set up chains
chains = harmonic.Chains(ndim)
chains.add_chains_3d(samples, lnprob)
chains_train, chains_infer = \
    harmonic.utils.split_data(chains, training_prop)

# Fit model
model = harmonic.model.KernelDensityEstimate(ndim, domain,
                                             hyper_parameters)
model.fit(chains_train.samples, chains_train.ln_posterior)

# Compute evidence
ev = harmonic.Evidence(chains_infer.nchains, model)
ev.add_chains(chains_infer)
ln_evidence, ln_evidence_std = ev.compute_ln_evidence()

\end{lstlisting}

%===============================================================================
\section{Conclusions}
\label{sec:conclusions}
%===============================================================================

We present the learnt harmonic mean estimator to solve the problematic large variance of the original estimator.  The construction of our estimator follows by interpreting the harmonic mean estimator as importance sampling and introducing a new target distribution that is learned to approximate the optimal but inaccessible target (the normalised posterior), while minimising the variance of the resulting estimator.  We discuss techniques to compute the variance of the estimator, its variance and to perform a number of additional computational sanity checks.  The estimator is implemented in the publicly available \texttt{harmonic} software code.  We demonstrate the application of our learnt harmonic mean estimator on numerous benchmark problems, including a number of pathological examples where the original harmonic mean estimator fails catastrophically.  In all cases our estimator is robust and highly accurate.  The current work opens up a number of avenues for future research.  For example, similar approaches can be taken in MCMC sampling more generally were appropriate target, sampling densities or proposal distributions may be learned.  Alternative more effective target models can be developed that better scale to higher dimensional settings.
%The estimator could also be combined with clustering approaches to compute local marginal likelihood estimates.
Since the learnt harmonic mean estimator is agnostic to the sampling strategy, it is also an ideal solution for computing the marginal likelihood for model comparison in simulation-based inference.  We are already actively pursuing this avenue of research, with promising preliminary results.

%===============================================================================
% Bayesian Statistics bibliography

%% ** The bibliograhy **
% \bibliographystyle{ba}
% \bibliography{<bib-data-file>}% place <bib-data-file>

% ** Acknowledgements **
% \begin{acknowledgement}
%   This work was supported by the Leverhulme Trust and by EPSRC grant EP/W007673/1.
%   For the purpose of open access, the authors have applied a Creative Commons
% Attribution (CC BY) licence to any Author Accepted Manuscript version arising.
% \end{acknowledgement}

%===============================================================================
% Springer Nature style

\backmatter

\bmhead{Acknowledgments}

This work was supported by the Leverhulme Trust and by EPSRC grant EP/W007673/1.
For the purpose of open access, the authors have applied a Creative Commons Attribution (CC BY) licence to any Author Accepted Manuscript version arising.

%===============================================================================
% Statistics & Computing bibliography

%\bibliographystyle{sn-basic}

% name your BibTeX data base
\bibliography{bib_journal_names_long,bib_myname,mybibs_new,bib,cosmo_references}

\end{document}